\documentclass[prd,superscriptaddress,nofootinbib,showpacs]{revtex4-2}

\usepackage{amsfonts,amssymb,amsmath}
\usepackage{mathrsfs}
\usepackage{xcolor}
\usepackage{bm}
\usepackage{hyperref}

\begin{document}

\title{Gauge-independent approach to inflation in quadratic gravity}

\author{Adrian Palomares}
\email{adri@tongji.edu.cn}
\affiliation{
	School of Physics Science and Engineering, Tongji University, Shanghai 200092, China
}

\author{Ying-li Zhang}
\email{yingli@tongji.edu.cn}
\affiliation{
	School of Physics Science and Engineering, Tongji University, Shanghai 200092, China
}
\affiliation{
	Institute for Advanced Study of Tongji University, Shanghai 200092, China
}
\affiliation{
	Kavli Institute for the Physics and Mathematics of the Universe (WPI), The University of Tokyo Institutes for Advanced Study, The University of Tokyo, Chiba 277-8583, Japan
}
\affiliation{
	Asia Pacific Center for Theoretical Physics, Pohang 37673, Korea
}
\affiliation{
	Center for Gravitation and Cosmology, Yangzhou University, Yangzhou 225009, China
}

\author{Jinsu Kim}
\email{kimjinsu@usst.edu.cn}
\thanks{Corresponding Author}
\affiliation{
    School of Physics, Faculty of Basic Sciences, University of Shanghai for Science and Technology, Shanghai 200093, China
}

\date{\today}

\begin{abstract}
	We investigate the scalar sector of linear cosmological perturbations in quadratic gravity. Working in the Einstein frame, we derive the equations of motion in a gauge-independent manner and express them in terms of three sets of gauge-invariant variables. This approach allows us to distinguish genuine physical effects from gauge artefacts, which is particularly relevant for assessing the stability of perturbations in this theory. In the superhorizon limit, we obtain the leading-order behaviour of the relevant gauge-invariant variables and analyse the perturbations in commonly used gauges. We find that the Newtonian gauge exhibits an apparent instability, characterised by the exponential growth of the metric perturbations. However, this growth is non-generic and gauge-dependent; in the other gauges analysed in this work, the perturbations remain well behaved within the perturbative regime. Our analysis also demonstrates how the evolution behaviour of a gauge-invariant variable changes under the frame transformation and clarifies the relation between results obtained in the Jordan and Einstein frames.
\end{abstract}

\maketitle

\section{\label{sec:intro} Introduction}
Inflation~\cite{Brout:1977ix,Starobinsky:1979ty,Starobinsky:1980te,Guth:1980zm,Linde:1981mu,Albrecht:1982wi} is a short period of accelerated expansion in the very early Universe that not only addresses the problems of the standard Big Bang framework, such as the horizon and flatness problems, but also seeds the primordial inhomogeneities through the quantum fluctuations of fields. After inflation, these fluctuations re-enter the Hubble horizon and leave measurable imprints in the cosmic microwave background (CMB) and the large-scale structure.

Cosmological perturbation theory provides a mathematical framework to study the dynamics of inflation and to connect theory with observations \cite{Kodama:1984ziu,Mukhanov:1990me,Malik:2008im}. It relies on the assumption that deviations from homogeneity and isotropy remain small, allowing for an expansion in perturbation variables around a homogeneous and isotropic background. It is therefore important to ensure that perturbations indeed remain within the perturbative regime. However, it should be stressed that the amplitude of perturbations is not a gauge-invariant statement; individual metric components may exhibit large or growing behaviour depending on the gauge choice, without implying a breakdown of the theory itself. Consequently, statements about the stability of solutions must ultimately rely on gauge-invariant quantities \cite{Bardeen:1980kt} or on physically meaningful observables such as the curvature perturbation $\mathcal{R}$.

According to the latest observational constraints from the CMB experiments such as Planck~\cite{Planck:2018jri}, BICEP/Keck~\cite{BICEP:2021xfz}, and ACT~\cite{ACT:2025fju,ACT:2025tim}, one of the most successful inflationary scenarios is the Starobinsky model~\cite{Starobinsky:1980te}. In this framework, the Einstein--Hilbert action is extended by a quadratic Ricci scalar $R^2$ correction that introduces a scalar degree of freedom commonly referred to as the scalaron. This provides a purely geometrical origin of inflation and, in the Einstein frame, is equivalent to a single-field inflationary model with a plateau-like potential along which the field slowly rolls. For a review, readers may refer to Ref.~\cite{DeFelice:2010aj}; see also Ref.~\cite{Nojiri:2010wj}. From a theoretical perspective, it is also well-motivated within the effective field theory (EFT) approach to gravity \cite{Donoghue:1994dn,Weinberg:2008hq}, in which higher-curvature corrections naturally arise as higher-dimensional operators suppressed by a cutoff scale.

Both the Starobinsky model and Einstein's theory of general relativity are expected to break down at sufficiently high energies, as neither provides an ultraviolet (UV) completion of gravity. From an EFT perspective, the gravitational action can be extended by including higher powers of $R$ as well as other curvature invariants. To the second order in the curvature, a more natural extension of Einstein's gravity theory would thus include $R^2$, $R_{\mu\nu}R^{\mu\nu}$, and $R_{\mu\nu\rho\sigma}R^{\mu\nu\rho\sigma}$. The resulting framework, known as quadratic gravity, was first systematically studied by Stelle \cite{Stelle:1976gc,Stelle:1977ry}, and it was shown that the inclusion of these additional quadratic terms renders the theory perturbatively renormalisable \cite{Fradkin:1981hx,Fradkin:1981iu,Avramidi:1985ki,Buchbinder:1992rb}. However, this improved UV behaviour comes at the cost of introducing additional propagating degrees of freedom. In addition to the massless graviton and the scalaron arising from the $R^2$ term, a massive spin-2 mode and an additional scalar mode come into play \cite{Stelle:1976gc,Stelle:1977ry}; see also Refs.~\cite{Deruelle:2010kf,Ivanov:2016hcm,Salvio:2018crh,DeFelice:2023psw,Hell:2023mph}.

At the classical level, the presence of higher derivatives is associated with the Ostrogradsky instability \cite{Ostrogradsky:1850fid}, which manifests itself through an unbounded Hamiltonian \cite{Motohashi:2014opa,Woodard:2015zca}. At the quantum level, the massive spin-2 state, which comes with the wrong sign for the kinetic term, can be quantised with a negative norm to avoid this instability \cite{Pais:1950za,Salvio:2015gsi,Salvio:2018crh}. While this raises concerns about unitarity, it is important to note that quadratic gravity can still be consistently interpreted as an EFT valid below a cutoff scale, or as a sector of a more complete theory in which these issues may be resolved. Various approaches to dealing with the ghost degree of freedom have been explored in the literature \cite{Salvio:2014soa,Einhorn:2014gfa,Anselmi:2017ygm,Anselmi:2018kgz,Anselmi:2020lpp,Bender:2007wu,Donoghue:2019fcb,Salvio:2020axm,Anselmi:2021rye,Edelstein:2021jyu,Kubo:2022dlx,Hell:2023rbf,Edelstein:2024jzu}, with different phenomenological interpretations depending on the physical regime considered.

In the context of inflationary cosmology, quadratic gravity has been previously studied at the level of linear perturbations \cite{Clunan:2009er,Nelson:2010wp,Deruelle:2010kf,Ivanov:2016hcm,Salvio:2017xul,DeFelice:2023psw,Bianchi:2025tyl}. It has been found that the tensor sector remains well behaved, with the additional massive spin-2 mode not spoiling the standard evolution of gravitational waves. Vector modes, which are absent in standard single-field inflation, become dynamical in this theory; however, their amplitude decreases as inflation proceeds, and they therefore do not lead to observable consequences.

In contrast, analyses of the scalar sector have led to differing interpretations. On the one hand, Ref.~\cite{Deruelle:2010kf}, where the analysis is done in the Einstein frame, concluded that the growth of certain perturbations observed in the Newtonian gauge is a gauge artefact rather than a physical instability. On the other hand, the more recent analysis of Ref.~\cite{DeFelice:2023psw}, where the analysis is performed in the Jordan frame, suggests that instabilities might persist even once the Newtonian gauge is avoided. It is important to note that these conclusions often rely on specific metric components or gauge-dependent variables, complicating their physical interpretation. Moreover, the different frame choices play a non-trivial role in comparing the two analyses.

This situation motivates a careful reanalysis of the scalar sector in a fully gauge-independent manner. Such an approach allows us to isolate the physical degrees of freedom and to determine whether any instability leads to the breakdown of perturbation theory. The aim of the present work is precisely to address this issue and to clarify the behaviour of scalar perturbations.

\section{\label{sec:setup} Setup}
At quadratic order in the curvature, the gravitational action is not restricted to the $R^2$ term that appears in Starobinsky inflation~\cite{Starobinsky:1980te}. In general, it may also include the scalars $R_{\mu\nu}R^{\mu\nu}$ and $R_{\mu\nu\rho\sigma}R^{\mu\nu\rho\sigma}$, leading to the most general local action built from curvature invariants up to second order. The corresponding action reads
\begin{align}
	S = \frac{M_{\mathrm{P}}^2}{2}
	\int d^4x \, \sqrt{-g} \, \left[
	R
	+\alpha R^2 
	+\beta R_{\mu\nu}R^{\mu\nu} 
	+\gamma R_{\mu\nu\rho\sigma}R^{\mu\nu\rho\sigma}
	\right]
	\,,\label{eqn:StelleS}
\end{align}
where $\alpha$, $\beta$, and $\gamma$ are constant coefficients of mass dimension $-2$, and $M_{\rm P} \equiv 1/\sqrt{8\pi G}$ is the reduced Planck mass.

Using the fact that the Gauss--Bonnet term, which is given by
\begin{align}
	\mathcal{G} =
	R^2 - 4R_{\mu\nu}R^{\mu\nu} + R_{\mu\nu\rho\sigma}R^{\mu\nu\rho\sigma}
	\,,
\end{align}
is topological in four dimensions~\cite{Lanczos:1938sf,Lovelock:1971yv}, the Riemann tensor-squared term, \textit{i.e.}, the $\gamma$ term, can be eliminated in favour of $R^2$ and $R_{\mu\nu}R^{\mu\nu}$. The action can then be written as
\begin{align}
	S =
	\frac{M_{\mathrm{P}}^2}{2}
	\int d^4x \, \sqrt{-g} \, \left[
	R
	+(\alpha - \gamma) R^2 
	+(\beta + 4\gamma) R_{\mu\nu}R^{\mu\nu} 
	\right]
	\,.
\end{align}
Alternatively, using the identity relating the Weyl tensor-squared term to the other invariants, namely~\cite{Weyl:1918ib}
\begin{align}
	C_{\mu\nu\rho\sigma}C^{\mu\nu\rho\sigma} =
	\mathcal{G}
	-\frac{2}{3}R^2 + 2R_{\mu\nu}R^{\mu\nu}
	\,,
\end{align}
one may eliminate the Ricci tensor-squared term, arriving at the following action:
\begin{align}
	S =
	\frac{M_{\mathrm{P}}^2}{2}
	\int d^4x \, \sqrt{-g} \, \left[
	R
	+\lambda R^2
	-\frac{\omega}{2} C_{\mu\nu\rho\sigma}C^{\mu\nu\rho\sigma}
	\right]
	\,,\label{eqn:JordanSWeyl}
\end{align} 
where the topological Gauss--Bonnet term has been ignored, and the coefficients $\lambda$ and $\omega$ are given by
\begin{align}
    \lambda =
	\alpha + \frac{1}{3}(\beta + \gamma)
    \,,\qquad
    \omega =
    -\left(
    \beta + 4\gamma
    \right)
    \,.\label{eqn:lambda-omega}
\end{align}
The action \eqref{eqn:JordanSWeyl} has recently been studied in Ref.~\cite{DeFelice:2023psw} in the context of inflationary cosmology. The parameters $\lambda$ and $\omega$ are restricted to positive values in the current work so as to avoid tachyonic instabilities in the spectrum. Readers may refer to Ref.~\cite{DeFelice:2023psw} for a more detailed discussion; see also \textit{e.g.}, Refs.~\cite{Deruelle:2010kf,Ivanov:2016hcm}.

It is well known that $f(R)$ gravity is equivalent to the scalar--tensor theory; see, for instance, Ref.~\cite{DeFelice:2010aj} for a review. As such, the $R^2$ term in the action \eqref{eqn:JordanSWeyl}, provided that its coefficient $\lambda$ is non-zero, can be mapped onto a scalar degree of freedom via the introduction of an auxiliary field $\varphi$. The action then reads
\begin{align}
	S =
	\frac{M_{\mathrm{P}}^2}{2}
	\int d^4x \, \sqrt{-g} \, \left[
	R + 2\lambda \varphi R - \lambda\varphi^2
	-\frac{\omega}{2}C_{\mu\nu\rho\sigma}C^{\mu\nu\rho\sigma}
	\right]
	\,.\label{eqn:JFaction}
\end{align}
One can easily see that varying the action with respect to the field $\varphi$ gives the constraint equation $\varphi = R$, given that $\lambda \neq 0$. Substituting $\varphi = R$ then recovers the original action \eqref{eqn:JordanSWeyl}. Via the Weyl rescaling, also commonly known as the conformal transformation,
\begin{align}
	g_{\mu\nu} \to g^\mathrm{E}_{\mu\nu} = f(\varphi) g_{\mu\nu}
	\,,\label{eqn:CT}
\end{align}
where we have used the index ``E'' to denote Einstein-frame quantities, and $f(\varphi) = 1 + 2\lambda \varphi$, the action \eqref{eqn:JFaction} can be brought to the Einstein frame as follows:
\begin{align}
	S = \int d^4x \, \sqrt{-g^\mathrm{E}} \, \left[
	\frac{M_{\mathrm{P}}^2}{2}R^\mathrm{E}
	-\frac{1}{2}g_\mathrm{E}^{\mu\nu}
	\partial_\mu \phi^{\rm E}      \partial_\nu \phi^{\rm E}
	-V^\mathrm{E}
	-\frac{\omega M^2_{\rm P}}{4}
	C^\mathrm{E}_{\mu\nu\rho\sigma}C^{\mu\nu\rho\sigma}_\mathrm{E}
	\right]
	\,.\label{eqn:EFaction}
\end{align}
Here, $\phi^\mathrm{E}$ is the canonically normalised field, which is given by
\begin{align}
	\phi^\mathrm{E} \equiv
	\sqrt{\frac{3}{2}}M_{\mathrm{P}} \ln f
	\,,
\end{align}
and $V^\mathrm{E}$ is the Einstein-frame potential, which is given by
\begin{align}
	V^\mathrm{E} =
	\frac{M_{\mathrm{P}}^2}{8\lambda} \left[
	1-\exp\left(
		-\sqrt{\frac{2}{3}}\frac{\phi^\mathrm{E}}{M_{\mathrm{P}}}
	\right)
	\right]^2
	\,.
\end{align}
We note that the Weyl-squared term is invariant under the Weyl rescaling. As we shall work extensively in the Einstein frame, we shall drop the index ``E'' from now on for brevity, unless otherwise stated.
The Einstein-frame action \eqref{eqn:EFaction} has been studied in Refs.~\cite{Deruelle:2010kf,Ivanov:2016hcm} in the context of inflationary cosmology.

\subsection{Background equations}
We consider a spatially flat Friedmann--Lema\^{\i}tre--Robertson--Walker (FLRW) background which, in the Einstein frame, is described by the metric
\begin{align}
    ds^2 = -dt^2 + a^2(t) \delta_{ij} dx^i dx^j
    \,.\label{eqn:FLRWbkg-metric}
\end{align}
Since the Weyl-squared term in the action \eqref{eqn:EFaction}, namely the $\omega$ term, vanishes at the background level, the background equations are identical to those of the standard single-field inflation model. In particular, the background equations of motion are given by the Friedmann equation and the Klein--Gordon equation,
\begin{align}
	3 M_{\mathrm{P}}^2 H^2 =
	\frac{\dot{\bar{\phi}}^2}{2}+V(\bar{\phi})
	\,,\qquad
	0 =
	\ddot{\bar{\phi}} + 3 H \dot{\bar{\phi}} + V'(\bar{\phi})
	\,,\label{eqn:bkgeom}
\end{align} 
where the dot indicates the derivative with respect to cosmic time $t$, and $H \equiv \dot{a}/a$ is the Hubble parameter. Here, we have used the notation $\bar{\phi}$ to denote that it is the background part of the inflaton field that depends only on time, $\bar{\phi} = \bar{\phi}(t)$.

\subsection{Perturbed equations}
Let us now consider linear perturbations around the background. Varying the action~\eqref{eqn:EFaction} with respect to $\delta g^{\mu\nu}$, we obtain
\begin{align}
	0 &= \left(
	R_{\mu\nu}-\frac{1}{2}g_{\mu\nu}R
	\right)
	-\frac{1}{M_{\mathrm{P}}^2}\left(\nabla_\mu\phi\nabla_\nu\phi-\frac{1}{2}g_{\mu\nu}\nabla_\rho\phi\nabla^\rho\phi
	-g_{\mu\nu}V\right)
	\nonumber\\ &\quad
	-\omega\bigg(
	\frac{1}{6}g_{\mu\nu}R^2
	-\frac{1}{2}g_{\mu\nu}R_{\rho\sigma}R^{\rho\sigma}
	-\frac{2}{3}g_{\mu\nu}\nabla^2R
	+g_{\mu\nu}\nabla_\rho\nabla_\sigma R^{\rho\sigma}
	+2R_{\mu}{}^\rho R_{\nu\rho}
	-\frac{2}{3}RR_{\mu\nu}
	\nonumber\\ &\quad\quad
	+\nabla^2 R_{\mu\nu}
	+\frac{2}{3}\nabla_\nu\nabla_\mu R
	-\nabla_\rho\nabla_\mu R_\nu{}^\rho
	-\nabla_\rho\nabla_\nu R_\mu{}^\rho
	\bigg)
	\,.\label{eqn:perteom-general}
\end{align}
The structure of the equations of motion \eqref{eqn:perteom-general} can be interpreted as the usual Einstein tensor $G_{\mu\nu}\equiv R_{\mu\nu}-g_{\mu\nu}R/2$ and the energy--momentum tensor of the scalar inflaton field $T_{\mu\nu}\equiv \nabla_{\mu}\phi\nabla_{\nu}\phi-(1/2)g_{\mu\nu}\nabla_{\rho}\phi\nabla^{\rho}\phi-g_{\mu\nu}V$, together with the contribution coming from the Weyl-squared term. One can check in a straightforward manner that the terms multiplied by $\omega$, \textit{i.e.}, the extra contribution coming from the Weyl-squared term, vanish on the FLRW background, as we mentioned earlier. Therefore, the Weyl-squared term only affects the perturbation dynamics.\footnote{
	An alternative, but equivalent, expression for the equations of motion is provided in Ref.~\cite{Deruelle:2010kf} as $0 = G_{\mu\nu} - T_{\mu\nu} / M_{\mathrm{P}}^2 - \omega B_{\mu\nu}$, where $B_{\mu\nu} \equiv 2 \nabla^{\alpha} \nabla^{\beta} C_{\mu\alpha\nu\beta} + G^{\alpha\beta} C_{\mu\alpha\nu\beta}$ is the Bach tensor~\cite{Bach:1921zdq}.
}

Considering only the scalar sector, we introduce perturbations of the metric as
\begin{align}
    ds^2 =
    -(1 + 2A) dt^2
    +2 a \partial_i B dt dx^i
    +a^2\left[
    (1 + 2\psi) \delta_{ij}
    -2 \partial_i \partial_j E
    \right] dx^i dx^j
    \,,\label{eqn:pertmetric}
\end{align}
while the scalar field is perturbed as
\begin{align}
    \phi(t,\mathbf{x})\equiv
    \bar{\phi}(t) + \delta\phi(t,\mathbf{x})
    \,,\label{eqn:pertinflaton}
\end{align}
where $\bar{\phi}(t)$ is the background part of the inflaton.
Substituting Eqs.~\eqref{eqn:pertmetric} and \eqref{eqn:pertinflaton} into Eq.~\eqref{eqn:perteom-general}, we find
\begin{align}
	0 &=
	6 H \dot{\psi}
	-6 H^2 A
	-2 H \triangle \dot{E}
	-2 H \triangle \left( \frac{B}{a} \right)
	-2 \triangle \frac{\psi}{a^2}
	-\frac{1}{M^2_{\rm P}} \left[
	\dot{\bar{\phi}} \delta\dot{\phi}
	-A \dot{\bar{\phi}}^2
	+V'(\bar{\phi}) \delta\phi
	\right]
	\nonumber\\ &\quad
	-\frac{2\omega}{3a^4} \triangle \triangle \left(
	A - \psi
	+a^2 \ddot{E}
	+a^2 H \dot{E}
	+a \dot{B}
	\right)
	\,,\label{eqn:Constraint-Hamilton}
\end{align}
from the $00$-component,
\begin{align}
	0 &=
	\partial_i\left\{
	-2 \dot{\psi}
	+2 H A
	-\frac{\dot{\bar{\phi}}}{M^2_{\rm P}} \delta\phi
	\right.
	\nonumber\\ &\quad
	\left.
	-\frac{2\omega}{3a^2} \triangle \left[
	\dot{A}
	-\dot{\psi}
	+a^2 \dddot{E}
	+3 a^2 H \ddot{E}
	+a^2 \left(2 H^2 + \dot{H}\right) \dot{E}
	+a \ddot{B}
	+a H \dot{B}
	\right]
	\right\}
	\,,\label{eqn:Constraint-Momentum}
\end{align}
from the $0i$-component, and
\begin{align}
	0 &=
	\left(
	\partial_i \partial_j
	-\frac{1}{3} \triangle \delta_{ij}
	\right)
	\left\{
	-\frac{\psi}{a^2}
	-\frac{A}{a^2}
	-\ddot{E}
	-3 H \dot{E}
	-\frac{\dot{B}}{a}
	-2 H \frac{B}{a}
	-\omega \left[
	\frac{\ddot{A}}{a^2}
	-\frac{\ddot{\psi}}{a^2}
	+H \left(
	\frac{\dot{A}}{a^2}
	-\frac{\dot{\psi}}{a^2}
	\right)
	\right.\right.
	\nonumber\\ &\quad
	\left.\left.
	-\frac{\triangle}{3a^4} \left(
	A - \psi
	\right)
	+\ddddot{E}
	+6 H \dddot{E}
	+\left(4 \dot{H} + 11 H^2\right) \ddot{E}
	+\left(
	\ddot{H} + 7 H \dot{H} + 6 H^3
	\right) \dot{E}
	\right. \right.
	\nonumber\\ &\quad
	\left. \left.
	+\frac{\dddot{B}}{a}
	+3 H \frac{\ddot{B}}{a}
	+\left(2 H^2 + \dot{H}\right) \frac{\dot{B}}{a}
	-\frac{\triangle}{3a^2} \left(
	\ddot{E}+H\dot{E}+\frac{\dot{B}}{a}
	\right)
	\right]
	\right\}
	\,,\label{eqn:Constraint-traceless}
\end{align}
from the spatial traceless component. Here, $\triangle \equiv \delta^{ij}\partial_i \partial_j$.

At this stage, one may fix a gauge and study the perturbation dynamics in the chosen gauge. Alternatively, one may construct gauge-invariant variables and examine the system in a gauge-independent manner. In this work, we adopt the latter approach.
Under a coordinate transformation $\tilde{x}_{\mu} = x_{\mu} + \lambda_{\mu}$ with $\lambda_i = \partial_i\lambda$, the scalar sector perturbations transform as follows:
\begin{align}
    A &\rightarrow
    \tilde{A} = A - \dot{\lambda}_0
    \,,\label{eqn:transf-A}\\
    B &\rightarrow 
    \tilde{B} = B - a\dot{\lambda} + \frac{\lambda_0}{a}
    \,,\label{eqn:transf-B}\\
    E &\rightarrow
    \tilde{E} = E + \lambda
    \,,\\
    \psi &\rightarrow
    \tilde{\psi} = \psi - H \lambda_0
    \,,\label{eqn:transf-psi}\\
    \delta\phi &\rightarrow
    \delta\tilde{\phi} = \delta\phi - \dot{\bar{\phi}} \lambda_0
    \,.\label{eqn:transf-phi}
\end{align}
These relations can be combined to construct gauge-invariant variables such as
\begin{align}
    \Psi &=
    \psi + a^2 H \left( \dot{E} + \frac{B}{a} \right)
    \,,&
    \Phi &= 
    A + \frac{d}{dt} \left[
    a^2 \left( \dot{E} + \frac{B}{a} \right)
    \right]
    \,,\nonumber\\
    \mathcal{R} &=
    \psi - \frac{H}{\dot{\bar{\phi}}} \delta\phi
    \,,&
    \chi &=
    a^2 \left(\frac{B}{a} + \dot{E} \right) + \frac{\delta\phi}{\dot{\bar{\phi}}}
    \,,\nonumber\\ \mathcal{A} &=
    A - \frac{d}{dt}\left(\frac{\delta \phi}{\dot{\bar{\phi}}}\right)
    \,,&
    \mathcal{A}&_{\psi}=
    A-\frac{d}{dt}\left(\frac{\psi}{H}\right)
    \,.\label{eqn:GIvars}
\end{align}
Expressing Eqs.~\eqref{eqn:Constraint-Hamilton}--\eqref{eqn:Constraint-traceless} in terms of the gauge-invariant variables $\Psi$, $\Phi$, and $\chi$, for instance, leads to 
\begin{align}
	0 &=
	3H(H\Phi-\dot{\Psi})
	-H^2\epsilon_1\Phi
	+\triangle\frac{\Psi}{a^2}
	+H^2\epsilon_1(\dot{\chi}-3H\chi)
	+\frac{\omega}{3a^4}\triangle\triangle\left(\Phi-\Psi\right)
	\,,\label{eqn:GIconstraint-Hamiltonian}\\
	0 &=
	\partial_i \left\{
	H\Phi
	-\dot{\Psi}
	-H^2\epsilon_1\chi
	-\frac{\omega}{3a^2}\triangle\left(\dot{\Phi}-\dot{\Psi}\right)
	\right\}
	\,,\label{eqn:GIconstraint-Momentum}\\
	0 &=
	\left(
	\partial_i\partial_i-\frac{1}{3}\delta_{ij}\triangle
	\right) \left\{
	\Psi+\Phi+\omega\left[\left(\ddot{\Phi}-\ddot{\Psi}\right)+H\left(\dot{\Phi}-\dot{\Psi}\right)-\frac{\triangle}{3a^2}\left(\Phi-\Psi\right)\right]
	\right\}
	\,,\label{eqn:GIconstraint-traceless}
\end{align}
where we have introduced the first Hubble slow-roll parameter $\epsilon_1$ defined as
\begin{align}
	\epsilon_1 \equiv
	-\frac{\dot{H}}{H^2}
	\,,
\end{align}
which characterises the deviation from the exact de Sitter (dS) background. We note that Eqs.~\eqref{eqn:GIconstraint-Hamiltonian}--\eqref{eqn:GIconstraint-traceless} recover the expressions given in Ref.~\cite{Deruelle:2010kf}.

One may also write the perturbed equations of motion in terms of $\mathcal{A}$, $\mathcal{R}$, and $\chi$ as follows:
\begin{align}
    0 &=
    3H(H\mathcal{A}-\dot{\mathcal{R}})
    +\frac{\triangle}{a^2}\left(\mathcal{R}+H\chi\right)
    -H^2\epsilon_1\mathcal{A}
    +\frac{\omega}{3a^4}\triangle \triangle\left[
    \mathcal{A}-\mathcal{R}+\dot{\chi}-H\chi
    \right]
    \,,\label{eqn:GIconstraint-Hamiltonian-ARsigma}\\
    0 &=
    \partial_i\left\{
    H\mathcal{A}
    -\dot{\mathcal{R}}
    +\frac{\omega}{3a^2}\triangle\left[
    \dot{\mathcal{R}}-\dot{\mathcal{A}}-\ddot{\chi}+H\dot{\chi}-H^2\epsilon_1\chi
    \right]
    \right\}
    \,,\label{eqn:GIconstraint-Momentum-ARsigma}\\
    0 &=
    \left(
    \partial_i\partial_j-\frac{1}{3}\delta_{ij}\triangle
    \right) \left\{
    -(\mathcal{R}+\mathcal{A})
    -\dot{\chi}
    -H\chi
    +\omega\left[
    \ddot{\mathcal{R}}
    -\ddot{\mathcal{A}}
    +H(\dot{\mathcal{R}}-\dot{\mathcal{A}})
    -\frac{\triangle}{3a^2}(\mathcal{R}-\mathcal{A})
    \right.\right.
    \nonumber\\&\quad
    \left.\left. 
    -\dddot{\chi}
    +\left(1-2\epsilon_1\right)H^2\dot{\chi}
    +\frac{\triangle}{3a^2}\dot{\chi}
    -(1-2\epsilon_1+\epsilon_2)H^3\epsilon_1\chi
    -\frac{H}{3a^2}\triangle\chi
    \right]
    \right\}
    \,,\label{eqn:GIconstraint-traceless-ARsigma}
\end{align}
where
\begin{align}
	\epsilon_{n>1} \equiv
	\frac{\dot{\epsilon}_{n-1}}{H\epsilon_{n-1}}
	\,.
\end{align}
In terms of $\mathcal{A}_{\psi}$, $\mathcal{R}$, and $\Psi$, instead, the perturbed equations of motion are given by
\begin{align}
    0 &=
    (3-\epsilon_1)H^2\mathcal{A}_{\psi}
    +(3-\epsilon_1)\epsilon_1 H^2 \mathcal{R}
    -H\epsilon_1\dot{\mathcal{R}}
    +\frac{\triangle}{a^2}\Psi
    +\frac{\omega}{3a^4}\triangle\triangle\left[
    \mathcal{A}_{\psi}
    -(1-\epsilon_1)\Psi
    +\frac{\dot{\Psi}}{H}
    \right]
    \,,\label{eqn:GIconstraint-Hamiltonian-ARpsi}\\
    0 &=
    \partial_i\left\{
    H(\mathcal{A}_{\psi}+\epsilon_1\mathcal{R})
    -\frac{\omega}{3a^2}\triangle\left[
    \dot{\mathcal{A}}_{\psi}
    -(1-2\epsilon_1)\dot{\Psi}
    +\frac{\ddot{\Psi}}{H}
    +H\epsilon_1\epsilon_2\Psi
    \right]
    \right\}
    \,,\label{eqn:GIconstraint-Momentum-ARpsi}\\
    0 &=
    \left(
    \partial_i\partial_j-\frac{1}{3}\delta_{ij}\triangle
    \right) \left\{
    \mathcal{A}_{\psi}
    +(1+\epsilon_1)\Psi
    +\frac{\dot{\Psi}}{H}
    +\omega \ddot{\mathcal{A}}_{\psi}
    +\omega H\dot{\mathcal{A}}_{\psi}
    -\frac{\omega}{3a^2}\triangle\left[
    \mathcal{A}_{\psi}-(1-\epsilon_1)\Psi
    \right]
    \right.
    \nonumber\\&\quad
    \left.
    +\frac{\omega \dddot{\Psi}}{H}
    +3\omega\epsilon_1\ddot{\Psi}
    -\omega H(1-2\epsilon_1-3\epsilon_1\epsilon_2)\dot{\Psi}
    -\frac{\omega}{3a^2}\triangle\left(\frac{\Psi}{H}\right)
    +\omega H^2\epsilon_1\epsilon_2(1-\epsilon_1+\epsilon_2+\epsilon_3)\Psi
    \right\}
    \,.\label{eqn:GIconstraint-traceless-ARpsi}
\end{align}

\section{\label{sec:GIanalysis} Evolution of gauge-invariant variables}
From the three systems of perturbed equations of motion obtained in the previous section, namely Eqs.~\eqref{eqn:GIconstraint-Hamiltonian}--\eqref{eqn:GIconstraint-traceless}, Eqs.~\eqref{eqn:GIconstraint-Hamiltonian-ARsigma}--\eqref{eqn:GIconstraint-traceless-ARsigma}, and Eqs.~\eqref{eqn:GIconstraint-Hamiltonian-ARpsi}--\eqref{eqn:GIconstraint-traceless-ARpsi}, one can derive a closed evolution equation for each of the gauge-invariant variables. In this section, we present the resultant expressions for such evolution equations. We then analyse the dynamical behaviour of the gauge-invariant variables by solving the evolution equations in the dS limit and taking the superhorizon limit. Meanwhile, the subhorizon analysis has been carried out in Ref.~\cite{DeFelice:2023psw}, with which we find agreement. The detailed derivation steps for the evolution equations of all the gauge-invariant variables, both in the superhorizon and subhorizon limits, can be found in Appendix~\ref{apdx:evolution-eqs}.

It is convenient to perform the calculation in Fourier space, \textit{i.e.},
\begin{align}
    X(t,\mathbf{x}) =
    \int d^3k \,
    X_\mathbf{k}(t) \,
    e^{i\mathbf{k}\cdot\mathbf{x}}
    \,,
\end{align}
where $X$ represents the gauge-invariant variable, $X = \{\Psi,\Phi,\mathcal{R},\chi,\mathcal{A},\mathcal{A}_\psi\}$. For simplicity, the wavenumber subscript ``$\mathbf{k}$'' will be dropped, $X_\mathbf{k} \rightarrow X$.

\subsection{Evolution of $\Psi$ and $\Phi$}
From Eqs.~\eqref{eqn:GIconstraint-Hamiltonian}--\eqref{eqn:GIconstraint-traceless}, we obtain fourth-order evolution equations for the gauge-invariant variables $\Psi$ and $\Phi$ as
\begin{align}
	0 &=
	\ddddot{\Phi}+H\alpha_3\dddot{\Phi}+H^2\alpha_2\ddot{\Phi}+H^3\alpha_1\dot{\Phi}+H^4\alpha_0\Phi
	\,,\label{eqn:EE-Phi}\\
    0 &=
    \ddddot{\Psi}+H\beta_3\dddot{\Psi}+H^2\beta_2\ddot{\Psi}+H^3\beta_1\dot{\Psi}+H^4\beta_0\Psi
    \,.\label{eqn:EE-Psi}
\end{align}
In Appendix~\ref{apdx:evolution-eqs}, we provide more details of the intermediate steps leading to the evolution equations, as well as the coefficients $\alpha_i$ and $\beta_i$.
In the superhorizon limit where $k \ll aH$, and to the zeroth order in the slow-roll parameters $\epsilon_i$ ($i=1,2,\cdots$), \textit{i.e.}, in the dS limit, Eqs.~\eqref{eqn:EE-Phi} and \eqref{eqn:EE-Psi} become
\begin{align}
	0 &=
	\ddddot{\Phi}+\left(\frac{1}{H^2\omega}-1\right)H^2\ddot{\Phi}+\frac{H}{\omega}\dot{\Phi}
	\,,\\
    0 &=
    \ddddot{\Psi}+\left(\frac{1}{H^2\omega}-1\right)H^2\ddot{\Psi}+\frac{H}{\omega}\dot{\Psi}
    \,.
\end{align}
Both $\Phi$ and $\Psi$ obey identical evolution equations at this order of approximation. Using the ansatz $\Phi,\Psi \propto e^{xHt}$ and solving the resulting algebraic equation for $x$, we obtain 
\begin{align}
    \Phi,\Psi &=
    C^{\Phi,\Psi}_1
    +C^{\Phi,\Psi}_2 e^{-Ht}
    +K^{\Phi,\Psi}_+e^{\frac{1}{2}(1+\theta)Ht}+K^{\Phi,\Psi}_-e^{\frac{1}{2}(1-\theta)Ht}
    \,,\label{eqn:sol-PhiPsi}
\end{align}
where
\begin{align}
	\theta&\equiv\sqrt{1-\frac{4}{H^2\omega}}
	\,,\label{eqn:theta}
\end{align}
and $C^{\Phi,\Psi}_1$, $C^{\Phi,\Psi}_2$, $K_+^{\Phi,\Psi}$, and $K_-^{\Phi,\Psi}$ are constants of integration for $\Phi$ and $\Psi$.

From Eq.~\eqref{eqn:theta}, it is evident that $\omega = \omega_\mathrm{crit} \equiv 4/H^2$ is a critical value to determine whether $\theta$ is real or pure imaginary. In either case, however, the last two terms in the solution~\eqref{eqn:sol-PhiPsi} always grow exponentially in time, causing the instability problem, as pointed out in Refs.~\cite{Deruelle:2010kf,Ivanov:2016hcm}; see also Ref.~\cite{DeFelice:2023psw} for the analysis in the Jordan frame.

\subsection{Evolution of $\mathcal{R}$}
We now turn to the gauge-invariant variable $\mathcal{R}$, whose evolution equation takes the form
\begin{align}
    0 &=
    \ddddot{\mathcal{R}}+H\gamma_3\dddot{\mathcal{R}}+H^2\gamma_2\ddot{\mathcal{R}}+H^3\gamma_1\dot{\mathcal{R}}+H^4\gamma_0\mathcal{R}
    \,.\label{eqn:R}
\end{align}
The derivation steps and expressions for the coefficients $\gamma_i$ up to $\mathcal{O}(\epsilon_i)$ are given in Appendix~\ref{apdx:evolution-eqs}.
In the superhorizon limit, Eq.~\eqref{eqn:R} reduces to
\begin{align}
    0 &=
    \ddddot{\mathcal{R}}+6H\dddot{\mathcal{R}}+\left(11+\frac{1}{\omega H^2}\right)H^2\ddot{\mathcal{R}}+\left(6+\frac{3}{H^2\omega}\right)H^3\dot{\mathcal{R}}
    \,,
\end{align}
up to the zeroth order in the slow-roll parameters $\epsilon_i$. Similarly to the previous case of $\Phi$ and $\Psi$, with the ansatz $\mathcal{R}\sim e^{xHt}$, the leading-order solution is found as
\begin{align}
    \mathcal{R} =
    C_1^{\mathcal{R}}
    +C_2^{\mathcal{R}}e^{-3Ht}
    +K_+^{\mathcal{R}}e^{-\frac{3}{2}\left(1+\frac{\theta}{3}\right)Ht}
    +K_-^{\mathcal{R}}
    e^{-\frac{3}{2}\left(1-\frac{\theta}{3}\right)Ht}
    \,,\label{eqn:sol-R}
\end{align} 
where $\theta$ is defined in Eq.~\eqref{eqn:theta}, and $C^\mathcal{R}_1$, $C^\mathcal{R}_2$, $K_+^\mathcal{R}$, and $K_-^{\mathcal{R}}$ are integration constants.
The first two terms correspond to the standard slow-roll solution, $\mathcal{R}_\mathrm{SR} \sim C + D/a^3$, consisting of a conserved mode and a decaying mode on superhorizon scales, while the additional two modes come from the Weyl-squared contribution.

For $\omega < \omega_\mathrm{crit} = 4/H^2$, $\theta$ is pure imaginary and the additional modes exhibit oscillating behaviour with exponentially decreasing amplitude. For $\omega > \omega_\mathrm{crit} = 4/H^2$, $\theta$ is real, and the additional modes decay exponentially. In either case, the extra modes coming from the Weyl-squared term decay with time, implying that $\mathcal{R}$ approaches a constant value on superhorizon scales, as in standard inflation. In other words, the gauge-invariant curvature perturbation $\mathcal{R}$, which is directly related to the CMB observables, freezes outside the Hubble horizon. This conclusion agrees with the result reported in Refs.~\cite{Starobinsky:2001xq,Deruelle:2010kf,Ivanov:2016hcm,DeFelice:2023psw}.

\subsection{Evolution of $\chi$}
The gauge-invariant variable $\chi$ satisfies
\begin{align}
    0 =
    \ddddot{\chi}+H\nu_3\dddot{\chi}+H^2\nu_2\ddot{\chi}+H^3\nu_1\dot{\chi}+H^4\nu_0\chi
    \,.
\end{align} 
The derivation steps and expressions for the coefficients $\nu_i$ up to $\mathcal{O}(\epsilon_i)$ are given in Appendix~\ref{apdx:evolution-eqs}.
In the superhorizon limit, together with the dS limit, the evolution equation for $\chi$ reads
\begin{align}
    0 =
    \ddddot{\chi}+2H\dddot{\chi}+\left(-3+\frac{1}{H^2\omega}\right)H^2\ddot{\chi}+\frac{3H}{\omega}\dot{\chi}\,.
\end{align}
Hence, the superhorizon solution is given by
\begin{align}
    \chi=
    C^{\chi}_1
    +C_2^{\chi}e^{-3Ht}
    +K_+^{\chi}
    e^{\frac{1}{2}\left(1+\theta\right)Ht}
    +K_-^{\chi}
    e^{\frac{1}{2}\left(1-\theta\right)Ht} 
    \,,
\end{align} 
where $\theta$ is defined in Eq.~\eqref{eqn:theta}, and $C^\chi_1$, $C^\chi_2$, $K_+^\chi$, and $K_-^{\chi}$ are integration constants.

The dependence on $\omega$ is the same as in the $\Psi$ and $\Phi$ cases. The last two modes, introduced by the higher-derivative contributions associated with the Weyl-squared term, suffer from the exponential instability on superhorizon scales.

\subsection{Evolution of $\mathcal{A}$}
The evolution equation for $\mathcal{A}$ is given by
\begin{align}
    0 =
    \ddddot{\mathcal{A}}+H\mu_3\dddot{\mathcal{A}}+H^2\mu_2\ddot{\mathcal{A}}+H^3\mu_1\dot{\mathcal{A}}+H^4\mu_0\mathcal{A}
    \,.
\end{align}
The derivation steps and exact expressions for the coefficients $\mu_i$ are given in Appendix~\ref{apdx:evolution-eqs}.
Taking the same limit as before, namely the superhorizon limit and the dS limit, leads to the following equation:
\begin{align}
    0 =
    \ddddot{\mathcal{A}}+12H\dddot{\mathcal{A}}+\left(53+\frac{1}{H^2\omega}\right)H^2\ddot{\mathcal{A}}+\left(102+\frac{5}{H^2\omega}\right)H^3\dot{\mathcal{A}}+\left(72+\frac{6}{H^2\omega}\right)H^4\mathcal{A}
    \,.
\end{align}
The solution is given by
\begin{align}
    \mathcal{A} =
    C_1^{\mathcal{A}}e^{-2Ht}
    +C_2^{\mathcal{A}}e^{-3Ht}
    +K_+^{\mathcal{A}}
    e^{-\frac{7}{2}\left(1+\frac{\theta}{7}\right)Ht} 
    +K_-^{\mathcal{A}}
    e^{-\frac{7}{2}\left(1-\frac{\theta}{7}\right)Ht} 
    \,,\label{eqn:sol-A}
\end{align}
where $\theta$ is defined in Eq.~\eqref{eqn:theta}, and $C^\mathcal{A}_1$, $C^\mathcal{A}_2$, $K_+^\mathcal{A}$, and $K_-^{\mathcal{A}}$ are integration constants.

In the case of $\mathcal{A}$, one can see that the dependence on $\omega$ is the same as for $\mathcal{R}$. We conclude that all the modes of $\mathcal{A}$ decay at the leading order.

\subsection{Evolution of $\mathcal{A}_{\psi}$}
Finally, let us examine the evolution equation for $\mathcal{A}_{\psi}$,
\begin{align}
    0 = 
    \ddddot{\mathcal{A}}_{\psi}+H\rho_3\dddot{\mathcal{A}}_{\psi}+H^2\rho_2\ddot{\mathcal{A}}_{\psi}+H^3\rho_1\dot{\mathcal{A}}_{\psi}+H^4\rho_0\mathcal{A}_{\psi}
    \,.
\end{align}
The derivation steps and expressions for the coefficients $\rho_i$ up to $\mathcal{O}(\epsilon_i)$ are given in Appendix~\ref{apdx:evolution-eqs}.
In the superhorizon limit, together with the dS limit, the evolution equation for $\mathcal{A}_\psi$ becomes
\begin{align}
    0 =
    \ddddot{\mathcal{A}}_{\psi}+6H\dddot{\mathcal{A}}_{\psi}+\left(11+\frac{1}{H^2\omega}\right)H^2\ddot{\mathcal{A}}_{\psi}+\left(6+\frac{3}{H^2\omega}\right)H^3\dot{\mathcal{A}}_{\psi}
    \,.
\end{align}
The solution is then given by
\begin{align}
    \mathcal{A}_{\psi} =
    C_1^{\mathcal{A}_{\psi}}
    +C_2^{\mathcal{A}_{\psi}}e^{-3Ht}+
    K_+^{\mathcal{A}_{\psi}}
    e^{-\frac{3}{2}\left(1+\frac{\theta}{3}\right)Ht}
    +K_-^{\mathcal{A}_{\psi}}
    e^{-\frac{3}{2}\left(1-\frac{\theta}{3}\right)Ht}
    \,,\label{eqn:sol-Apsi}
\end{align}
where $\theta$ is defined in Eq.~\eqref{eqn:theta}, and $C^{\mathcal{A}_{\psi}}_1$, $C^{\mathcal{A}_{\psi}}_2$, $K_+^{\mathcal{A}_{\psi}}$, and $K_-^{\mathcal{A}_{\psi}}$ are integration constants.

We note that the $\mathcal{A}_\psi$ solution \eqref{eqn:sol-Apsi} takes the same form as the $\mathcal{R}$ solution \eqref{eqn:sol-R}. In other words, it has three decaying modes and one constant mode.
This outcome, which is found in a gauge-independent way, agrees with the analysis of Ref.~\cite{Deruelle:2010kf}. However, it differs from the result of Ref.~\cite{DeFelice:2023psw}, where a linearly growing solution is reported instead of the constant mode.

In Ref.~\cite{DeFelice:2023psw}, the linearly growing mode is found in the flat gauge to take the form
\begin{align}
	\mathcal{A}_{\psi}^{\rm DF} \approx 
	c_4 \left(
	1 + \frac{t}{18\lambda H}
	\right)
	\,,\label{eqn:sol-Apsi-DF}
\end{align}
where $\lambda$, which corresponds to $\beta$ in the notation of Ref.~\cite{DeFelice:2023psw}, is given in Eq.~\eqref{eqn:lambda-omega}, $c_4$ is the integration constant, and we have put the superscript ``DF'' to indicate that it is the solution reported in Ref.~\cite{DeFelice:2023psw}.
It is important to note that the analysis of Ref.~\cite{DeFelice:2023psw} is performed in the Jordan frame, where the action takes the form of Eq.~\eqref{eqn:JordanSWeyl}. Therefore, $H$, $t$, and $\mathcal{A}_\psi^{\rm DF}$ are actually the Jordan-frame quantities.
Furthermore, the second term in Eq.~\eqref{eqn:sol-Apsi-DF} is of first order in the slow-roll parameters; in the Jordan frame, the first Hubble slow-roll parameter is approximately given by $\epsilon_1^{\rm J} \approx 1/(36 \lambda H^2)$. To the zeroth order in the slow-roll parameters, the solution \eqref{eqn:sol-Apsi-DF} would thus correspond to a constant mode.

In our case, where the analysis is performed in the Einstein frame, the constant solution is the outcome of the usage of the leading-order contributions in the slow-roll parameters. As shown in Eq.~\eqref{eqn:rho0}, the leading order for $\rho_0$ is already $\mathcal{O}(\epsilon_2)$. If we take into account the contributions from terms of first order in the slow-roll parameters, the solution that corresponds to the constant solution $\mathcal{A}_\psi = C_1^{\mathcal{A}_\psi}$ becomes
\begin{align}
	\mathcal{A}_\psi =
	C_1^{\mathcal{A}_\psi} e^{\epsilon_2 H t} \approx
	C_1^{\mathcal{A}_\psi} \left(
	1 + \epsilon_2 H t
	\right)
	\,.\label{eqn:sol-Apsi-NLO}
\end{align}
We stress again that $\mathcal{A}_\psi$, $H$, $\epsilon_2$, and $t$ in Eq.~\eqref{eqn:sol-Apsi-NLO} are the Einstein-frame quantities. This next-to-leading order matches the linear-growth behaviour found in Ref.~\cite{DeFelice:2023psw}. Note that the linear growth, being suppressed by slow-roll parameters in both the Einstein and Jordan frames, is harmless to perturbation theory; see also Ref.~\cite{DeFelice:2023psw}.
We can also show via Weyl rescaling that the solutions, Eq.~\eqref{eqn:sol-Apsi-DF} and Eq.~\eqref{eqn:sol-Apsi-NLO}, are, in fact, equivalent. The detailed derivation is presented in Appendix~\ref{apdx:WeylRescaling}.

It is also worth mentioning that if we use the relation,
\begin{align}
	\mathcal{A}_\psi =
	\mathcal{A}
	-\frac{d}{dt}\left(
	\frac{\mathcal{R}}{H}
	\right)
	\,,\label{eqn:Apsi-A-R}
\end{align}
which comes from the definitions of $\mathcal{A}_\psi$ and $\mathcal{A}$, substituting the solutions \eqref{eqn:sol-R} and \eqref{eqn:sol-A} would lead to the incorrect result of $\mathcal{A}_\psi \propto e^{-3 H t / 2}$. This is because the approximated solutions do not carry the information about the constraints between the variables. This is where the gauge-independent approach proves useful, as no variables are constructed from other variables; each evolution equation for the gauge-invariant variables is worked out from the system of equations consisting of gauge-invariant variables.

\section{\label{sec:gauges} Analysis of gauges}
As the analysis performed in Sec.~\ref{sec:GIanalysis} is carried out in terms of the gauge-invariant variables, the results are valid in any gauge. However, the analysis of the metric perturbations differs depending on the gauge choice. In this section, we analyse how the metric variables evolve in commonly used gauges.

Let us first summarise the leading-order behaviour of the gauge-invariant variables found in the previous section:
\begin{gather}
   \Phi,\Psi,\chi \propto 
   e^{\frac{1}{2}(1+\theta)Ht}
   \,,\quad
   \mathcal{R} \simeq \text{const.}
   \,,\quad
   \mathcal{A} \propto e^{-2Ht}
   \,,\quad
   \mathcal{A}_{\psi} \simeq \text{const.}
\end{gather}
Strictly speaking, the leading-order behaviour of $\Phi$, $\Psi$, and $\chi$ depends on the value of $\omega$. As long as $\omega < 4/H^2$, however, we have $\Phi,\Psi,\chi \propto e^{Ht/2}$. When $\omega > 4/H^2$, the growth rate ranges from $e^{Ht/2}$ to $e^{Ht}$, and our analysis presented in this section would correspond to the value of $\omega$ that leads to the weakest growth rate of the dominant mode. We stress that the exact dependence is irrelevant for the qualitative conclusions of our analysis.

\subsection*{Newtonian gauge}
The Newtonian gauge is defined by the conditions $E=0$ and $B=0$. The remaining metric variables $A$ and $\psi$ reduce to
\begin{align}
    \psi_\mathrm{N} =
    \Psi \propto e^{\frac{1}{2}Ht}
    \,,\quad
    A_\mathrm{N} =
    \Phi \propto e^{\frac{1}{2}Ht}
    \,.
\end{align}
It is also possible to read the scalar-field perturbation; it behaves as
\begin{align}
	\frac{\delta\phi_\mathrm{N}}{\dot{\bar{\phi}}} =
	\chi \propto e^{\frac{1}{2}Ht}
	\,.
\end{align}
As such, in the Newtonian gauge, all the perturbation variables grow exponentially with time. As a consequence, the fundamental assumption of perturbation theory that perturbations remain small compared to the background quantities is violated. In other words, the linear perturbation theory breaks down in this gauge. This implies that the Newtonian gauge is not suitable for analysing the perturbative dynamics in this framework.

\subsection*{Flat gauge}
The slicing of the flat gauge is fixed by the condition $\psi=0$. For the threading, \textit{i.e.}, the spatial coordinate transformation, one may choose either $B=0$ or $E=0$. In this gauge, the perturbations evolve as
\begin{align}
	\frac{\delta\phi_\mathrm{F}}{\dot{\bar{\phi}}} =
	-\frac{\mathcal{R}}{H} \simeq
	\text{const.}
	\,,\quad
	A_\mathrm{F} =
	\mathcal{A}_{\psi} \simeq \text{const.}
	\,,
\end{align}
and depending on the threading, we have either
\begin{align}
	E_\mathrm{F} =
	\int \frac{\Psi}{a^2H} \, dt \propto
	e^{-\frac{3}{2}Ht}
	\,,
\end{align}
for $B = 0$, or
\begin{align}
	B_\mathrm{F} =
	\frac{\Psi}{aH} \propto
	e^{-\frac{1}{2}Ht}
	\,,
\end{align}
for $E = 0$.\footnote{
	We note that $B$ and $E$ inherit the dependence on $\omega$ from $\Psi$, but remain bounded for all admissible values of $\omega$.
}
Note that $A_\mathrm{F}$ corresponds to the gauge-invariant variable $\mathcal{A}_\psi$ shown in Eq.~\eqref{eqn:Apsi-A-R}. All the perturbations in the flat gauge either decay or remain constant in time. Therefore, the linear perturbation theory remains valid.

\subsection*{Comoving gauge}
The slicing of the comoving gauge is fixed by the condition $\delta\phi=0$. For the threading, the literature typically adopts either $B=0$ or $E=0$. In this gauge, the metric perturbations are
\begin{align}
    \psi_\mathrm{C} =
    \mathcal{R} \simeq \text{const.}
    \,,\quad
    A_\mathrm{C} =
    \mathcal{A} \propto e^{-2Ht}
    \,,
\end{align}
and depending on the choice for the threading, one may have 
\begin{align}
    E_\mathrm{C} =
    \int \frac{\chi}{a^2} \, dt \propto
    e^{-\frac{3}{2}Ht}
    \,,
\end{align}
for $B = 0$, or
\begin{align}
	B_\mathrm{C} =
	\frac{\chi}{a} \propto
	e^{-\frac{1}{2}Ht}
	\,,
\end{align}
for $E = 0$.
The leading-order contributions are either constant or decaying with time. Consequently, the linear perturbation theory remains valid; the comoving slicing exhibits bounded metric perturbations.

In Ref.~\cite{DeFelice:2023psw}, working in the Jordan frame, the authors analysed the unitary gauge defined by $\delta R=0$ and $E=0$, which corresponds to $\delta \phi=0$ and $E=0$ in the Einstein frame. In particular, they showed that the gauge-invariant variable,
\begin{align}
	\mathcal{B} \equiv a(B + a\dot{E}) + \psi/H
	\,,
\end{align}
grows as $\mathcal{B}\propto e^{Ht/2}$. On this basis, it was argued that the growth of the metric component $g_{0i}=a\partial_i B$ signals an instability. In our analysis of the comoving slice, as $B_\mathrm{C} \propto e^{-Ht/2}$, we see that $g_{0i} = a \partial_i B_\mathrm{C} \propto e^{Ht/2}$ also features such a growth. Notably, the same behaviour arises in the flat gauge as well.
However, judging from the following three points, we argue that this growth does not lead to the breakdown of perturbation theory:
\begin{itemize}
	\item[(1)] The growth of $g_{0i}$ is a gauge effect.
	\item[(2)] The perturbativity is determined by $B$, not $\mathcal{B}$.
	\item[(3)] The causal structure is not affected by the growth of $g_{0i}$.
\end{itemize}
We present our arguments in detail as follows:\\

\underline{About the argument (1)}\\
The growth of $g_{0i}$ reflects a gauge effect rather than a physical instability. By performing an appropriate gauge transformation, \textit{e.g.}, imposing $B=0$, the growing behaviour in $g_{0i}$ is removed, while all the other scalar perturbations remain bounded, resulting in an isotropic solution. In particular, the curvature perturbation $\mathcal{R}$ remains constant on superhorizon scales. This indicates that the growth does not correspond to a dynamical degree of freedom.\\

\underline{About the argument (2)}\\
Even in gauges where $E=0$ is imposed, the growth of $\mathcal{B}$ does not imply the breakdown of perturbation theory. While $\mathcal{B} = a B \propto e^{Ht/2}$ appears to suggest an exponential instability, it is important to recall that $B \propto e^{-Ht/2}$ decays. The conclusion that the growth of $\mathcal{B}$ invalidates the inflationary FLRW background is therefore not a coordinate-invariant way of testing the perturbativity. A more appropriate framework is provided by projecting onto the orthonormal frame of comoving observers~\cite{Wald:1984rg} (see also Ref.~\cite{Mitsou:2019nhj}), where the basis is given by
\begin{align*}
	e_{\hat{0}} = \partial_t
	\,,\qquad
	e_{\hat{i}} = \frac{1}{a} \partial_i
	\,,
\end{align*}
where hatted indices label local Lorentz frame directions. The components of the metric in this frame of comoving observers are defined by $\hat{g}_{\hat{a} \hat{b}} \equiv g_{\mu\nu} e_{\hat{a}}^\mu e_{\hat{b}}^\nu$, where $e_{\hat{a}}^\mu$, extracted from $e_{\hat{a}} = e_{\hat{a}}^\mu \partial_\mu$, are, in our case, given by
\begin{align*}
	e_{\hat{0}}^\mu = \delta_{\hat{0}}^\mu
	\,,\qquad
	e_{\hat{i}}^\mu = \frac{1}{a} \delta_{\hat{i}}^\mu
	\,.
\end{align*}
The deviations from the background are then given by $\delta\hat{g}_{\hat{a} \hat{b}} \equiv \hat{g}_{\hat{a} \hat{b}} - \hat{\eta}_{\hat{a} \hat{b}}$, where $\hat{\eta}_{\hat{a} \hat{b}} = \bar{g}_{\mu\nu}e^\mu_{\hat{a}} e^\nu_{\hat{b}} = \mathrm{diag}(-1, +1, +1, +1)$ is the FLRW background translated to the Minkowski metric. The deviation of the metric component in question, \textit{i.e.}, $\delta\hat{g}_{\hat{0} \hat{i}}$ is thus
\begin{align*}
	\delta\hat{g}_{\hat{0} \hat{i}} = 
	a^{-1} \partial_i \mathcal{B} =
	\partial_i B
	\,.
\end{align*}
The physically relevant $0i$ component is therefore controlled by $\partial_i B$, rather than $\partial_i \mathcal{B}$. As $B$ decays, we see that the amplitude of $\delta\hat{g}_{\hat{0} \hat{i}} = \partial_i B \propto e^{-Ht/2}$ also decays, confirming the validity of perturbation theory.\\

\underline{About the argument (3)}\\
By examining whether the congruence defined by the worldlines of $dx^i=0$ remains timelike throughout the inflationary trajectory, one may show that the growing $g_{0i}$ in the present case does not lead to the violation of causality, which otherwise would make the coordinate system, \textit{i.e.}, the gauge choice, pathological \cite{Wald:1984rg}.\footnote{
	We thank Misao Sasaki for helping clarify this point.
}
Let us consider the ADM decomposition \cite{Arnowitt:1959ah,Arnowitt:1962hi,Witten1962Gravitation},
\begin{align}
	ds^2=-\alpha^2dt^2+\gamma_{ij}\left(dx^i+\beta^idt\right)\left(dx^j+\beta^jdt\right)
	\,,\label{eqn:ADM}
\end{align}
where $\alpha$ is the lapse function, $\beta^i$ is the shift vector, and $\gamma_{ij}$ is the spatial metric.
To leading order, we identify
\begin{align}
	\alpha_\mathrm{C} \simeq 1 + A_\mathrm{C}\,,\quad \beta_{i\mathrm{C}} = a \partial_i B_\mathrm{C}
	\,.
\end{align}
For the threading condition $E=0$, the spatial metric at linear order is given by
\begin{align}
	\gamma^{ij}_\mathrm{C} \simeq a^{-2}(1-2\psi_\mathrm{C})\delta^{ij}
	\,,
\end{align}
from which it follows that
\begin{align}
	\beta^i_\mathrm{C} \simeq a^{-1} \delta^{ij} \partial_j B_\mathrm{C}
	\,.
\end{align}
We see from Eq.~\eqref{eqn:ADM} that, to leading order,
\begin{align}
	g_{00\mathrm{C}}=
	-\alpha^2_\mathrm{C}
	+\beta_{i\mathrm{C}}\beta^i_\mathrm{C}
	<0
	\,.
\end{align}
Therefore, the comoving slice, despite exhibiting a growing $g_{0i}$, preserves the timelike character of the congruence of worldlines $dx^i=0$, and hence, it does not break the causal structure and defines a regular coordinate system. Following the same reasoning, one may also arrive at the same conclusion for the flat gauge.

\vspace{0.5cm}

In summary, the apparent exponential instability of the Newtonian gauge is absent in all the other gauges considered; the flat and comoving gauges are well-behaved coordinate systems. This observation confirms that the instability seen in the Newtonian gauge is a gauge artefact rather than a physical effect.

This contrasting behaviour may be understood from the structure of the perturbed equations of motion.\footnote{
	We thank Guillem Dom\`{e}nech for a remark on this analysis.
}
Let us consider first Eqs.~\eqref{eqn:GIconstraint-Hamiltonian}--\eqref{eqn:GIconstraint-traceless}, written in terms of $\Psi$, $\Phi$, and $\chi$, which in the Newtonian gauge correspond to the metric and field perturbations $\psi$, $A$, and $\delta\phi$, respectively. In the Hamiltonian and momentum constraints, Eqs.~\eqref{eqn:GIconstraint-Hamiltonian} and \eqref{eqn:GIconstraint-Momentum}, the contribution from the Weyl-squared term enters through gradient terms and is thus suppressed in the superhorizon limit $k \ll a H$. However, as these two constraints couple all three variables, the last equation~\eqref{eqn:GIconstraint-traceless} is required to close the system, and in this equation, the Weyl-squared term is not gradient-suppressed. As a result, $\Psi$, $\Phi$ and $\chi$ all exhibit the exponential growth.

In contrast, the remaining two sets of equations, Eqs.~\eqref{eqn:GIconstraint-Hamiltonian-ARsigma}--\eqref{eqn:GIconstraint-traceless-ARsigma} and Eqs.~\eqref{eqn:GIconstraint-Hamiltonian-ARpsi}--\eqref{eqn:GIconstraint-traceless-ARpsi}, are structurally different. Here, the third variable, namely $\chi$ and $\Psi$, respectively, enters the Hamiltonian and momentum constraints through gradient terms. Therefore, in the superhorizon limit, to leading order in the gradient expansion, these two constraints form a closed system for the bounded variables, $\mathcal{A}$ and $\mathcal{R}$ in the first set, and $\mathcal{A}_\psi$ and $\mathcal{R}$ in the second, without any Weyl-squared sourcing. Hence, they remain well behaved. The Weyl-squared term survives only in the last equation of each set, which determine the remaining variable, $\chi$ and $\Psi$; this is consistent with the solutions found in Sec.~\ref{sec:GIanalysis}.

The analysis of the structure of the perturbed equations makes the origin of the apparent instability clear. The Newtonian gauge is the only gauge considered here in which the perturbation variables are directly sourced by the Weyl-squared term; in all the other gauges this sourcing is gradient-suppressed.

\section{\label{sec:conc} Conclusion}
In this work, we have investigated the stability of scalar perturbations during inflation in quadratic gravity with a Weyl-squared term. Working in the Einstein frame, we have adopted a gauge-independent approach based on gauge-invariant variables, which allows us to distinguish genuine physical effects from gauge artefacts.

We have derived the equations of motion for the scalar perturbations and expressed them explicitly in terms of the gauge-invariant variables. The resulting evolution equations have then been solved in the superhorizon limit and in the de Sitter limit. This analysis has allowed us to determine the leading-order time dependence of the relevant gauge-invariant variables. We have shown that while some of these variables exhibit exponentially growing behaviour, the physically relevant variables remain well behaved. In particular, the curvature perturbation $\mathcal{R}$ approaches a constant on superhorizon scales, consistent with the behaviour expected in standard single-field inflationary scenarios.

To better understand the implications for the scalar perturbations, we have subsequently analysed commonly used gauge choices. In the Newtonian gauge, we found that the scalar perturbations grow exponentially with time. Consequently, they may eventually become comparable to the background quantities, apparently invalidating the perturbative expansion in this gauge. On the other hand, no such exponential instability has been observed in at least two other gauges, namely the flat and comoving gauges. We have thus identified the exponential instability observed in the Newtonian gauge as a gauge artefact rather than a physical instability.

Our results are consistent with those of Ref.~\cite{Deruelle:2010kf}. The Weyl-squared term in quadratic gravity does not make the inflationary background unviable, provided that a well-defined gauge is chosen. Additionally, our work extends the analysis of Ref.~\cite{Deruelle:2010kf} by working systematically with multiple sets of gauge-invariant variables, which clarifies the role of the constraints between them.

We have also shown that our Einstein-frame analysis, once the frame difference is properly taken into account, agrees with the Jordan-frame analysis performed in Ref.~\cite{DeFelice:2023psw}, which reopened the question of the phenomenological viability of quadratic gravity. However, our physical interpretation of the results differs from that of Ref.~\cite{DeFelice:2023psw}. In the comoving gauge, for instance, we have emphasised that the apparent growth of the metric component $g_{0i}$ does not lead to the breakdown of cosmological perturbation theory. This conclusion rests on the facts that the metric component remains well-behaved in the perturbative regime once the frame of comoving observers is taken and that the causal structure remains unbroken. Furthermore, as we have demonstrated, the scalar perturbations remain bounded. In particular, the curvature perturbation $\mathcal{R}$ approaches a constant on superhorizon scales.

We stress that our analysis is restricted to linear perturbations around the flat FLRW background in the superhorizon limit with the de Sitter approximation. A number of important questions therefore remain open. First, it would be meaningful to extend the present analysis beyond the leading-order slow-roll approximation in order to assess how the growing modes behave in a realistic inflationary background where the Hubble parameter varies in time. Second, a detailed computation of the cosmological observables, such as the power spectrum, the spectral index, and the tensor-to-scalar ratio, in the presence of the Weyl-squared term would allow for a direct comparison with current CMB observational data. Finally, our analysis does not address the fate of the massive spin-2 ghost mode. We leave these important questions for future work.

\begin{acknowledgments}
	We thank Guillem Dom\`{e}nech, Cristiano Germani, Misao Sasaki, and Anna Tokareva for discussions.
	AP and YZ are supported by the Fundamental Research Funds for the Central Universities, and by the Project 12475060 supported by NSFC, Project 24ZR1472400 sponsored by Natural Science Foundation of Shanghai, and Shanghai Pujiang Program 24PJA134.
	JK is supported by National Natural Science Foundation of China (NSFC) under Grant No. 12505079.
\end{acknowledgments}

\appendix
\section{\label{apdx:evolution-eqs} Evolution equations}

\subsection*{Evolution equation for $\Phi$}
The evolution equation for $\Phi$ can be obtained through the following steps:
\begin{enumerate}
	\item From Eq.~\eqref{eqn:GIconstraint-Momentum}, we find $\chi$.
	\item Taking the time derivative of $\chi$, we find $\dot{\chi}$.
	\item We substitute $\chi$ and $\dot{\chi}$ into Eq.~\eqref{eqn:GIconstraint-Hamiltonian} and solve for $\ddot{\Psi}$.
	\item We then plug $\ddot{\Psi}$ into Eq.~\eqref{eqn:GIconstraint-traceless} and find the solution for $\dot{\Psi}$.
	\item Taking the time derivative of this equation and using the results of $\ddot{\Psi}$ and $\dot{\Psi}$, we isolate $\Psi$ in terms of $\Phi$ and its derivatives.
	\item After obtaining the solution for $\Psi$, we again take the time derivative and use the results for $\dot{\Psi}$ and $\Psi$.
\end{enumerate}
Following these steps would result in the evolution equation of $\Phi$,
\begin{align}
    0 =
    \ddddot{\Phi}+H\alpha_3\dddot{\Phi}+H^2\alpha_2\ddot{\Phi}+H^3\alpha_1\dot{\Phi}+H^4\alpha_0\Phi
    \,.
\end{align}
As the exact expressions for the coefficients $\alpha_i$ are rather lengthy, we present them in the subhorizon and superhorizon limits, up to the order unity in $aH/k$ and $k/(aH)$, respectively. Furthermore, we take terms up to the first order in the slow-roll parameters. Then, the coefficients are given by
\begin{align}
	\alpha_3 &\approx
	8 + 2 \epsilon_1 - \epsilon_2
	\,,\\
	\alpha_2 &\approx
	\frac{2 k^2}{a^2 H^2} + 5 + \frac{1}{H^2 \omega}
	+24 \epsilon_1 - 9 \epsilon_2
	\,,\\
	\alpha_1 &\approx
	\frac{k^2}{a^2 H^2} \left(
	4 + 2 \epsilon_1 - \epsilon_2
	\right)
	-74 - \frac{13}{H^2 \omega}
	+113 \epsilon_1 - 29 \epsilon_2
	+\frac{2 \epsilon_1}{H^2 \omega} - \frac{\epsilon_2}{H^2 \omega}
	\,,\\
	\alpha_0 &\approx
	\frac{k^4}{a^4 H^4}
	+\frac{k^2}{a^2 H^2}\left(
	-16 + \frac{1}{H^2 \omega} + \frac{70}{3} \epsilon_1 - 4 \epsilon_2
	\right)
	\nonumber\\&\quad
	+204 + \frac{159}{H^2 \omega}
	-380 \epsilon_1 + 153 \epsilon_2
	-\frac{70 \epsilon_1}{H^2 \omega}
	+\frac{21 \epsilon_2}{2 H^2 \omega}
	\,,
\end{align}
in the subhorizon limit, and
\begin{align}
   \alpha_3 &\approx
   2 \epsilon_1 - \epsilon_2
   \,,\\
   \alpha_2 &\approx
   -1 + \frac{1}{H^2 \omega} + 2 \epsilon_1
   \,,\\
   \alpha_1 &\approx
   \frac{1}{H^2 \omega}
   +3 \epsilon_1 + \epsilon_2
   +\frac{2 \epsilon_1}{H^2 \omega}
   -\frac{\epsilon_2}{H^2 \omega}
   \,,\\
   \alpha_0 &\approx
   -\frac{\epsilon_2}{H^2 \omega}
   \,,
\end{align}
in the superhorizon limit.

\subsection*{Evolution equation for $\Psi$}
The evolution equation for $\Psi$ can be obtained through the following steps:
\begin{enumerate}
	\item From Eq.~\eqref{eqn:GIconstraint-Momentum}, we find $\chi$.
	\item Taking the time derivative of $\chi$, we find $\dot{\chi}$.
	\item We substitute $\chi$ and $\dot{\chi}$ into Eq.~\eqref{eqn:GIconstraint-Hamiltonian} and solve for $\ddot{\Phi}$.
    \item Substituting $\ddot{\Phi}$ into Eq.~\eqref{eqn:GIconstraint-traceless}, we find $\dot{\Phi}$.
    \item We then take the time derivative of the resulting expression and use the solutions of $\ddot{\Phi}$ and $\dot{\Phi}$ to isolate $\Phi$.
    \item Finally, we take one more time derivative and use the solutions for $\dot{\Phi}$ and $\Phi$.
\end{enumerate}
Following these steps would result in the evolution equation of $\Psi$,
\begin{align}
    0 =
    \ddddot{\Psi}+H\beta_3\dddot{\Psi}+H^2\beta_2\ddot{\Psi}+H^3\beta_1\dot{\Psi}+H^4\beta_0\Psi
    \,.
\end{align}
As the exact expressions for the coefficients $\beta_i$ are lengthy, we present them in the subhorizon and superhorizon limits, up to the order unity in $aH/k$ and $k/(aH)$, respectively. Up to the first order in the slow-roll parameters, they are given by
\begin{align}
	\beta_3 &\approx
	8 + 2\epsilon_1 - \epsilon_2
	\,,\\
	\beta_2 &\approx 
	\frac{2 k^2}{a^2 H^2} + 5 + 24 \epsilon_1 - 9 \epsilon_2 + \frac{1}{H^2 \omega}
	\,,\\
	\beta_1 &\approx
	\frac{k^2}{a^2 H^2} \left(
	4 + 2 \epsilon_1 - \epsilon_2
	\right)
	- 74 + \frac{11}{H^2 \omega}
	+ 113 \epsilon_1 - 29 \epsilon_2
	+\frac{2 \epsilon_1}{H^2 \omega}
	-\frac{\epsilon_2}{H^2 \omega}
	\,,\\
	\beta_0 &\approx
	\frac{k^4}{a^4 H^4}
	+\frac{k^2}{a^2 H^2}\left(
	-16 + \frac{1}{H^2 \omega} + \frac{70 \epsilon_1}{3} - 4 \epsilon_2
	\right)
	\nonumber\\&\quad
	+204 - \frac{63}{H^2 \omega}
	-380 \epsilon_1 + 153 \epsilon_2
	+\frac{74 \epsilon_1}{H^2 \omega}
	-\frac{45 \epsilon_2}{2 H^2 \omega}
	\,,
\end{align}
in the subhorizon limit, and
\begin{align}
	\beta_3 &\approx
	4 \epsilon_1 - \epsilon_2
	\,,\\
	\beta_2 &\approx
	-1 + \frac{1}{H^2 \omega} + 2 \epsilon_1
	\,,\\
	\beta_1 &\approx
	\frac{1}{H^2 \omega}
	+\epsilon_1 + \epsilon_2
	+\frac{2 \epsilon_1}{H^2 \omega}
	-\frac{\epsilon_2}{H^2 \omega}
	\,,\\
	\beta_0 &\approx
	-\frac{\epsilon_2}{H^2 \omega}
	\,,
\end{align}
in the superhorizon limit.

\subsection*{Evolution equation for $\mathcal{R}$}
The evolution equation for $\mathcal{R}$ can be obtained through the following steps:
\begin{enumerate}
    \item From Eq.~\eqref{eqn:GIconstraint-Hamiltonian-ARsigma}, we isolate $\mathcal{A}$ and calculate $\dot{\mathcal{A}}$ and $\ddot{\mathcal{A}}$. 
    \item We then substitute them into Eq.~\eqref{eqn:GIconstraint-Momentum-ARsigma} and Eq.~\eqref{eqn:GIconstraint-traceless-ARsigma}.
    \item After the substitution, Eq.~\eqref{eqn:GIconstraint-Momentum-ARsigma} becomes $F(\dddot{\mathcal{R}}, \ddot{\mathcal{R}},\dot{\mathcal{R}},\mathcal{R},\dddot{\chi},\ddot{\chi},\dot{\chi},\chi)=0$ and Eq.~\eqref{eqn:GIconstraint-traceless-ARsigma} becomes $G( \ddot{\mathcal{R}},\dot{\mathcal{R}},\mathcal{R},\ddot{\chi},\dot{\chi},\chi)=0$; we find $\dddot{\chi}$ from the former and $\ddot{\chi}$ from the latter.
    \item We take the time derivative of $G( \ddot{\mathcal{R}},\dot{\mathcal{R}},\mathcal{R},\ddot{\chi},\dot{\chi},\chi)=0$, substitute $\dddot{\chi}$ and $\ddot{\chi}$ found in the previous step, and find $\dot{\chi}$ from the resultant equation.
    \item We take another time derivative, use the solutions for $\ddot{\chi}$ and $\dot{\chi}$, and find $\chi$.
    \item We repeat the process one last time using the solutions for $\dot{\chi}$ and $\chi$. 
\end{enumerate}
Following these steps would result in the evolution equation of $\mathcal{R}$,
\begin{align}
    0 = 
    \ddddot{\mathcal{R}}+H\gamma_3\dddot{\mathcal{R}}+H^2\gamma_2\ddot{\mathcal{R}}+H^3\gamma_1\dot{\mathcal{R}}+H^4\gamma_0\mathcal{R}
    \,.
\end{align}
As the exact expressions for the coefficients $\gamma_i$ are lengthy, we present them in the subhorizon and superhorizon limits, up to the order unity in $aH/k$ and $k/(aH)$, respectively. Up to the first order in the slow-roll parameters, they are given by
\begin{align}
	\gamma_3 &\approx
	8 + 2\epsilon_1 + \epsilon_2
	\,,\\
	\gamma_2 &\approx
	2\left(\frac{k}{aH}\right)^2
    +19 + \frac{2}{H^2\omega}
    +4\epsilon_1+5\epsilon_2
    +\frac{4\epsilon_1}{3H^2\omega}
    -\frac{\epsilon_2}{2H^2\omega}
	\,,\\
	\gamma_1 &\approx
	\left(\frac{k}{aH}\right)^2
	(4+2\epsilon_1+\epsilon_2)
	+12
	+\frac{5}{H^2\omega}
	-\frac{1}{2H^4\omega^2}
	\nonumber\\&\quad
	-3\epsilon_1
	+4\epsilon_2
	+\frac{2\epsilon_1}{H^2\omega}
	-\frac{5\epsilon_1}{6H^4\omega^2}
	+\frac{\epsilon_2}{2H^4\omega^2}
	\,,\\
	\gamma_0 &\approx
	\left(\frac{k}{aH}\right)^4
	+\left(\frac{k}{aH}\right)^2\left(
	-2
	+\frac{2}{H^2\omega}
	+\frac{10\epsilon_1}{3}
	+\frac{4\epsilon_1}{3H^2\omega}
	-\frac{\epsilon_2}{2H^2\omega}
	\right)
	\nonumber\\&\quad
	+\frac{1}{H^2\omega}
	-\frac{1}{4H^6\omega^3}
	-\frac{\epsilon_1}{3H^2\omega}
	+\frac{\epsilon_2}{2H^2\omega}
	-\frac{5\epsilon_1}{6H^4\omega^2}
	-\frac{\epsilon_2}{2H^4\omega^2}
	-\frac{\epsilon_1}{2H^6\omega^3}
	+\frac{3\epsilon_2}{8H^6\omega^3}
	\,,
\end{align}
in the subhorizon limit, and
\begin{align}
	\gamma_3 &\approx
	6 + \epsilon_2
	+\frac{4 \epsilon_1 H^2 \omega}{1 + 2 H^2 \omega}
	\,,\\
	\gamma_2 &\approx
	11 + \frac{1}{H^2 \omega} + 2 \epsilon_1 + 3\epsilon_2
	-\frac{6\epsilon_1}{1+2H^2\omega}
	\,,\\
	\gamma_1 &\approx
	6+\frac{3}{H^2 \omega}
	-3\epsilon_1
	+2\epsilon_2
	+\frac{\epsilon_2}{H^2\omega}
	\,,\\
	\gamma_0& \approx
	0
	\,,
\end{align}
in the superhorizon limit.

\subsection*{Evolution equation for $\chi$}
The evolution equation for $\chi$ can be obtained through the following steps:
\begin{enumerate}
	\item From Eq.~\eqref{eqn:GIconstraint-Hamiltonian-ARsigma}, we isolate $\mathcal{A}$ and calculate $\dot{\mathcal{A}}$ and $\ddot{\mathcal{A}}$. 
	\item We then substitute them into Eq.~\eqref{eqn:GIconstraint-Momentum-ARsigma} and Eq.~\eqref{eqn:GIconstraint-traceless-ARsigma}.
	\item After the substitution, Eq.~\eqref{eqn:GIconstraint-Momentum-ARsigma} becomes $F(\dddot{\mathcal{R}}, \ddot{\mathcal{R}},\dot{\mathcal{R}},\mathcal{R},\dddot{\chi},\ddot{\chi},\dot{\chi},\chi)=0$ and Eq.~\eqref{eqn:GIconstraint-traceless-ARsigma} becomes $G( \ddot{\mathcal{R}},\dot{\mathcal{R}},\mathcal{R},\ddot{\chi},\dot{\chi},\chi)=0$; we find $\dddot{\mathcal{R}}$ from the former and $\ddot{\mathcal{R}}$ from the latter.
    \item We take the time derivative of $G( \ddot{\mathcal{R}},\dot{\mathcal{R}},\mathcal{R},\ddot{\chi},\dot{\chi},\chi)=0$ and substitute $\dddot{\mathcal{R}}$ and $\ddot{\mathcal{R}}$ found in the previous step, and find $\dot{\mathcal{R}}$ from the resultant equation.
    \item We take another time derivative, use the solutions for $\ddot{\mathcal{R}}$ and $\dot{\mathcal{R}}$, and find $\mathcal{R}$.
    \item We repeat the process one last time using the solutions for $\dot{\mathcal{R}}$ and $\mathcal{R}$.
\end{enumerate}
Following these steps would result in the evolution equation of $\chi$,
\begin{align}
    0 = \ddddot{\chi}+H\nu_3\dddot{\chi}+H^2\nu_2\ddot{\chi}+H^3\nu_1\dot{\chi}+H^4\nu_0\chi
    \,.
\end{align}
As the exact expressions for the coefficients $\nu_i$ are lengthy, we present them in the subhorizon and superhorizon limits, up to the order unity in $aH/k$ and $k/(aH)$, respectively. Up to the first order in the slow-roll parameters, they are given by
\begin{align}
	\nu_3 &\approx
	8 - 2\epsilon_1 + \epsilon_2
	\,,\\
	\nu_2 &\approx
	2\left(\frac{k}{aH}\right)^2
	+23
	+\frac{1}{H^2\omega}
	-24\epsilon_1
	+6\epsilon_2
	\,,\\
	\nu_1 &\approx
	\left(\frac{k}{aH}\right)^2 \left(
	4 - 2\epsilon_1 + \epsilon_2
	\right)
	-4
	+\frac{7}{H^2\omega}
	-\frac{7\epsilon_1}{3}
	+2\epsilon_2
	-\frac{2\epsilon_1}{H^2\omega}
	+\frac{\epsilon_2}{H^2\omega}
	\,,\\
	\nu_0 &\approx
	\left(\frac{k}{aH}\right)^4
	+\left(\frac{k}{aH}\right)^2 \left(
	2
	+\frac{1}{H^2\omega}
	-\frac{14\epsilon_1}{3}
	+\epsilon_2
	\right)
	-112
	+\frac{61}{H^2\omega}
	\nonumber\\&\quad
	+\frac{772\epsilon_1}{3}
	-50\epsilon_2
	-\frac{47\epsilon_1}{H^2\omega}
	+\frac{25\epsilon_2}{2H^2\omega}
	\,,
\end{align}
in the subhorizon limit, and
\begin{align}
	\nu_3 &\approx
	2
	+\frac{2\epsilon_1}{1+12H^2\omega}
	\left(\frac{k}{aH}\right)^{-2}
	+\epsilon_2
	-\frac{2\epsilon_1}{1+12H^2\omega}
	\,,\\
	\nu_2 &\approx
	-3
	+\frac{1}{H^2 \omega}
	-\frac{2\epsilon_1}{1+12H^2\omega}
	\left(\frac{k}{aH}\right)^{-2}
	+\frac{11\epsilon_1}{9}
	-\epsilon_2
	+\frac{25\epsilon_1}{9(1+12H^2\omega)}
	\,,\\
	\nu_1 &\approx
	\frac{36}{1+12H^2\omega}
	+\frac{3}{(1+12H^2\omega)H^2 \omega}
	+\frac{2 \epsilon_1}{(1+12H^2\omega)H^2\omega}\left(
	\frac{k}{aH}
	\right)^{-2}
	+\frac{280 H^2 \omega \epsilon_1}{3(1+12H^2\omega)}
	\nonumber\\&\quad
	+\frac{9\epsilon_1}{1+12H^2\omega}
	+\frac{12\epsilon_2}{1+12H^2\omega}
	-\frac{2\epsilon_1}{(1+12H^2\omega)H^2 \omega}
	+\frac{\epsilon_2}{(1+12H^2\omega)H^2 \omega}
	\,,\\
	\nu_0 &\approx
	-\frac{\epsilon_1}{H^2\omega}
	-\frac{12\epsilon_1}{1+12H^2\omega}
	\,,
\end{align}
in the superhorizon limit.

\subsection*{Evolution equation for $\mathcal{A}$}
The evolution equation for $\mathcal{A}$ can be obtained through the following steps:
\begin{enumerate}
    \item From Eq.~\eqref{eqn:GIconstraint-Momentum-ARsigma}, we find $\dot{\mathcal{R}}$ and compute $\ddot{\mathcal{R}}$.
    \item Taking the time derivative of Eq.~\eqref{eqn:GIconstraint-Momentum-ARsigma} and substituting $\ddot{\mathcal{R}}$ and $\dot{\mathcal{R}}$, we obtain the first equation that is independent of $\mathcal{R}$.
    \item We then substitute $\dot{\mathcal{R}}$ into Eq.~\eqref{eqn:GIconstraint-Hamiltonian-ARsigma} and find $\mathcal{R}$.
    \item Plugging $\mathcal{R}$, $\dot{\mathcal{R}}$, and $\ddot{\mathcal{R}}$ into Eq.~\eqref{eqn:GIconstraint-traceless-ARsigma}, we obtain the second equation that is independent of $\mathcal{R}$.
    \item From the first equation, we find $\dddot{\chi}$ and feed it to the second equation from which we can find $\ddot{\chi}$.
    \item We take the time derivative of the resulting equation and substitute $\dddot{\chi}$ and $\ddot{\chi}$ into it, from which we can find $\dot{\chi}$.
    \item We repeat the same process twice more.
\end{enumerate}
Following these steps would result in the evolution equation of $\mathcal{A}$,
\begin{align}
    0 =
    \ddddot{\mathcal{A}}+H\mu_3\dddot{\mathcal{A}}+H^2\mu_2\ddot{\mathcal{A}}+H^3\mu_1\dot{\mathcal{A}}+H^4\mu_0\mathcal{A}
    \,.
\end{align}
The coefficients $\mu_i$ are given, without taking any limit, as follows:
\begin{align}
    \mu_3 &=
    12 - 2\epsilon_1 + \epsilon_2
    \,,\\
    \mu_2 &=
    53+\frac{1}{\omega H^2}-36\epsilon_1+9\epsilon_2+6\epsilon_1^2-9\epsilon_1\epsilon_2+3\epsilon_2\epsilon_3 +2\left(\frac{ k}{a H}\right)^2
    \,,\\
    \mu_1 &=
    102
    -159 \epsilon_1
    +26 \epsilon_2
    +78 \epsilon_1^2
    -75 \epsilon_1 \epsilon_2
    +18 \epsilon_2 \epsilon_3
    -12 \epsilon_1^3
    -9 \epsilon_1 \epsilon_2^2
    +30 \epsilon_1^2 \epsilon_2
    +3 \epsilon_2 \epsilon_3^2
    \nonumber\\&\quad
    -15 \epsilon_1 \epsilon_2 \epsilon_3
    +3 \epsilon_2 \epsilon_3 \epsilon_4
    +\frac{5}{H^2 \omega}
    -\frac{2 \epsilon_1}{H^2 \omega}
    +\frac{\epsilon_2}{H^2 \omega}
    +\left(
    8
    -2 \epsilon_1
    +\epsilon_2
    \right) \left(
    \frac{k}{aH}
    \right)^2
	\,,\\
    \mu_0 &=
    72
    -204 \epsilon_1
    +24 \epsilon_2
    +204 \epsilon_1^2
    -140 \epsilon_1 \epsilon_2
    +26 \epsilon_2 \epsilon_3
    -84 \epsilon_1^3
    +150 \epsilon_1^2 \epsilon_2
    -33 \epsilon_1 \epsilon_2^2
    \nonumber\\&\quad
    -57 \epsilon_1 \epsilon_2 \epsilon_3
    +9 \epsilon_2 \epsilon_3^2
    +9 \epsilon_2 \epsilon_3 \epsilon_4
    +12 \epsilon_1^4
    -42 \epsilon_1^3 \epsilon_2
    +25 \epsilon_1^2 \epsilon_2 \epsilon_3
    +29 \epsilon_1^2 \epsilon_2^2
    -3 \epsilon_1 \epsilon_2^3
    \nonumber\\&\quad
    -11 \epsilon_1 \epsilon_2^2 \epsilon_3
    -8 \epsilon_1 \epsilon_2 \epsilon_3^2
    -8 \epsilon_1 \epsilon_2 \epsilon_3 \epsilon_4
    +\epsilon_2 \epsilon_3^3
    +3 \epsilon_2 \epsilon_3^2 \epsilon_4
    +\epsilon_2 \epsilon_3 \epsilon_4^2
    +\epsilon_2 \epsilon_3 \epsilon_4 \epsilon_5
    \nonumber\\&\quad
    +\frac{6}{H^2 \omega}
    -\frac{8 \epsilon_1}{H^2 \omega}
    +\frac{2 \epsilon_2}{H^2 \omega}
    +\frac{2 \epsilon_1^2}{H^2 \omega}
    -\frac{3 \epsilon_1 \epsilon_2}{H^2 \omega}
    +\frac{\epsilon_2 \epsilon_3}{H^2 \omega}
    \nonumber\\&\quad    
    +\frac{1}{H^2\omega}\left(
    \frac{k}{aH}
    \right)^2
    -\frac{26 \epsilon_1}{3} \left(
    \frac{k}{aH}
    \right)^2
    +2 \epsilon_2 \left(
    \frac{k}{aH}
    \right)^2
    +2 \epsilon_1^2 \left(
    \frac{k}{aH}
    \right)^2
    \nonumber\\&\quad
    -3 \epsilon_1 \epsilon_2 \left(
    \frac{k}{aH}
    \right)^2
    +\epsilon_2 \epsilon_3 \left(
    \frac{k}{aH}
    \right)^2
    +8 \left(
    \frac{k}{aH}
    \right)^2
    +\frac{k^4}{a^4 H^4}
    \,.
\end{align}

\subsection*{Evolution equation for $\mathcal{A}_{\psi}$}
The evolution equation for $\mathcal{A}_\psi$ can be obtained through the following steps:
\begin{enumerate}
    \item From Eq.~\eqref{eqn:GIconstraint-traceless-ARpsi}, we find $\dddot{\Psi}$, and from Eq.~\eqref{eqn:GIconstraint-Momentum-ARpsi}, we find $\ddot{\Psi}$. 
    \item Taking the time derivative of Eq.~\eqref{eqn:GIconstraint-Momentum-ARpsi} and substituting the expressions for $\dddot{\Psi}$ and $\ddot{\Psi}$, we find $\dot{\Psi}$.
    \item Taking another time derivative and using the expressions for $\ddot{\Psi}$ and $\dot{\Psi}$, we find $\Psi$.
    \item Taking one last time derivative and using the expressions for $\dot{\Psi}$ and $\Psi$, we obtain one equation that is independent of $\Psi$.
    \item Substituting the expressions for $\Psi$ and its derivatives in Eq.~\eqref{eqn:GIconstraint-Hamiltonian-ARpsi} gives another equation that is independent of $\Psi$.
    \item Calling these two equations $F(\dddot{\mathcal{A}}_{\psi},\ddot{\mathcal{A}}_{\psi},\dot{\mathcal{A}}_{\psi},\mathcal{A}_{\psi},\dddot{\mathcal{R}},\ddot{\mathcal{R}},\dot{\mathcal{R}},\mathcal{R})=0$ and $G(\ddot{\mathcal{A}}_{\psi},\dot{\mathcal{A}}_{\psi},\mathcal{A}_{\psi},\ddot{\mathcal{R}},\allowbreak\dot{\mathcal{R}},\mathcal{R}) = 0$, we find $\dddot{\mathcal{R}}$ from $F$ and $\ddot{\mathcal{R}}$ from $G$.
    \item Taking the time derivative of $G$ and using the expressions for $\dddot{\mathcal{R}}$ and $\ddot{\mathcal{R}}$, we find $\dot{\mathcal{R}}$.
    \item Taking another derivative and using the expressions for $\ddot{\mathcal{R}}$ and $\dot{\mathcal{R}}$, we find $\mathcal{R}$.
    \item We take the time derivative once more and substitute the expressions for $\dot{\mathcal{R}}$ and $\mathcal{R}$ into it.
\end{enumerate}
Following these steps would result in the evolution equation of $\mathcal{A}_\psi$,
\begin{align}
    0 =
    \ddddot{\mathcal{A}}_{\psi}+H\rho_3\dddot{\mathcal{A}}_{\psi}+H^2\rho_2\ddot{\mathcal{A}}_{\psi}+H^3\rho_1\dot{\mathcal{A}}_{\psi}+H^4\rho_0\mathcal{A}_{\psi}
    \,.
\end{align}
As the exact expressions for the coefficients $\rho_i$ are lengthy, we present them in the subhorizon and superhorizon limits, up to the order unity in $aH/k$ and $k/(aH)$, respectively. Up to the first order in the slow-roll parameters, they are given by
\begin{align}
	\rho_3 &\approx
	12 - 2 \epsilon_1 - \epsilon_2
	\,,\\
	\rho_2 &\approx
	41 + \frac{1}{H^2 \omega}
	- 20 \epsilon_1 - 13 \epsilon_2
	+2\left(\frac{k}{aH}\right)^2
	\,,\\
	\rho_1 &\approx
	22 + \frac{11}{H^2\omega}
	-\frac{193\epsilon_1}{3}
	-49\epsilon_2
	-\frac{\epsilon_2}{H^2\omega}
	+8\left(\frac{k}{aH}\right)^2
	-2 \epsilon_1\left(\frac{k}{aH}\right)^2
	-\epsilon_2\left(\frac{k}{aH}\right)^2
	\,,\\
	\rho_0 &\approx
	84
	+\frac{9}{H^2\omega}
	-164\epsilon_1
	+37\epsilon_2
	+\frac{42\epsilon_1}{H^2\omega}
	-\frac{49\epsilon_2}{2H^2\omega}
	-4\left(\frac{k}{aH}\right)^2
	+\frac{1}{H^2\omega}\left(\frac{k}{aH}\right)^2
	\nonumber\\&\quad
	+\frac{22\epsilon_1}{3}\left(\frac{k}{aH}\right)^2
	-6\epsilon_2\left(\frac{k}{aH}\right)^2
	+\left(\frac{k}{aH}\right)^4
	\,,
\end{align}
in the subhorizon limit, and
\begin{align}
	\rho_3 &\approx
	6 + 2 \epsilon_1 - \epsilon_2
	-\frac{2 \epsilon_1}{1 + 2 H^2 \omega}
	-\frac{2 \epsilon_1}{1 + 4 H^2 \omega}
	\,,\\
	\rho_2 &\approx
	11 + \frac{1}{H^2 \omega} + 2 \epsilon_1 - 6 \epsilon_2
	-\frac{12 \epsilon_1}{1 + 2 H^2 \omega}
	-\frac{6 \epsilon_1}{1 + 4 H^2 \omega}
	\,,\\
	\rho_1 &\approx
	6 + \frac{3}{H^2 \omega} - 3 \epsilon_1 - 11 \epsilon_2
	-\frac{18 \epsilon_1}{1 + 2 H^2 \omega}
	+\frac{4 \epsilon_1}{1 + 4 H^2 \omega}
	-\frac{\epsilon_2}{H^2 \omega}
	\,,\\
	\rho_0 &\approx
	-6 \epsilon_2 - \frac{3 \epsilon_2}{H^2 \omega}
	\,,\label{eqn:rho0}
\end{align}
in the superhorizon limit.

\section{\label{apdx:WeylRescaling} Conformal transformation}
The conformal transformation, also known as the Weyl rescaling, allows one to go from one frame to another. For instance, through the Weyl rescaling, we can change the action \eqref{eqn:JFaction}, which we call the Jordan-frame action, to the Einstein-frame action \eqref{eqn:EFaction}. In this section, we first investigate relationships between Jordan-frame quantities and Einstein-frame quantities. We then discuss the behaviour of the gauge-invariant variable $\mathcal{A}_\psi$ in both frames. In the following discussion, we shall use the superscript J (E) to denote the Jordan (Einstein) frame.

Let us recall the conformal transformation we considered in Sec.~\ref{sec:setup},
\begin{align}
	g_{\mu\nu}^\mathrm{J} \to
	g_{\mu\nu}^\mathrm{E} =
	f(\varphi) g_{\mu\nu}^\mathrm{J}
	\,,
\end{align}
where $\varphi$ is the auxiliary field which is identified with the Ricci scalar in the Jordan frame, namely $\varphi = R^\mathrm{J}$. Comparing the line element in the Jordan frame to that in the Einstein frame at the background level, one may see that
\begin{align}
	dt^\mathrm{E} = \sqrt{f} dt^\mathrm{J}
	\,,\quad 
	a^\mathrm{E} = \sqrt{f} a^\mathrm{J} 
	\,,
\end{align}
where $t$ is cosmic time, and $a$ is the scale factor.
From this observation, we obtain the relation between the Jordan-frame Hubble parameter $H^\mathrm{J} \equiv (da^\mathrm{J}/dt^\mathrm{J}) / a^\mathrm{J}$ and the Einstein-frame Hubble parameter $H^\mathrm{E} \equiv (da^\mathrm{E}/dt^\mathrm{E}) / a^\mathrm{E}$ as follows:
\begin{align}
	H^\mathrm{E} =
	\frac{1}{\sqrt{f}} H^\mathrm{J}
	+\frac{f' \dot{\varphi}}{2 f^{3/2}}
	\,,
\end{align}
where $\dot{\varphi} \equiv d\varphi/dt^\mathrm{J}$.

At the level of the first-order perturbations, the comparison of the line element in different frames gives rise to
\begin{align}
	A^\mathrm{E} = 
	A^\mathrm{J} 
	+\frac{f' \delta\varphi}{2 f}
	\,,\quad
	\psi^\mathrm{E} = 
	\psi^\mathrm{J} 
	+\frac{f' \delta\varphi}{2 f}
	\,,\quad
	B^\mathrm{E} = B^\mathrm{J}
	\,,\quad
	E^\mathrm{E} = E^\mathrm{J}
	\,,\label{eqn:CT-metricpert}
\end{align}
where $f$ here is to be understood as the background part of $f$, and $\delta\varphi$ would correspond to the perturbation of the Ricci scalar in the Jordan frame, $\delta\varphi = \delta R^\mathrm{J}$.
We note that if $\delta R^\mathrm{J} = 0$, which is the condition for the unitary gauge, then the metric perturbation variables become frame-independent. Therefore, the analysis performed in this gauge would remain the same under the frame change. In general, however, the evolution of the perturbation variables depends on the frame choice. Furthermore, the choice of the gauge is frame-dependent. For instance, the Newtonian gauge in the Einstein frame is identical to the Newtonian gauge in the Jordan frame, while the flat gauge in the Jordan frame cannot be transferred to the flat gauge in the Einstein frame as $\psi^\mathrm{J} = 0$ and $\psi^\mathrm{E} = 0$ cannot be simultaneously satisfied.

To understand the discrepancy between the result of Ref.~\cite{DeFelice:2023psw} on $\mathcal{A}_\psi$, namely Eq.~\eqref{eqn:sol-Apsi-DF}, and our result on $\mathcal{A}_\psi$, namely Eq.~\eqref{eqn:sol-Apsi-NLO}, let us concentrate on the evolution of the gauge-invariant variable $\mathcal{A}_\psi$ in the Jordan-frame flat gauge.\footnote{
	We stress that the same analysis can be repeated in other gauge choices.
}
In the Jordan-frame flat gauge, where $\psi^\mathrm{J} = 0$ and $E=0$, as $\mathcal{A}^\mathrm{J}_\psi = A^\mathrm{J}$, $\delta\varphi$ is given by
\begin{align}
	\delta\varphi =
	-6 H^\mathrm{J} \dot{\mathcal{A}}^\mathrm{J}_\psi
	-24 (H^\mathrm{J})^2 \mathcal{A}^\mathrm{J}_{\psi}
	-12 \dot{H}^\mathrm{J} \mathcal{A}^\mathrm{J}_{\psi}
	+\frac{2k^2}{(a^\mathrm{J})^2}\left(
	\mathcal{A}^\mathrm{J}_{\psi} 
	+a^\mathrm{J} \dot{B}^\mathrm{J} 
	+3 a^\mathrm{J} H^\mathrm{J} B^\mathrm{J}
	\right)
	\,.
\end{align}
From Eq.~\eqref{eqn:CT-metricpert}, we find
\begin{align}
	\mathcal{A}^\mathrm{E}_\psi &=
	A^\mathrm{E} - \frac{d}{dt^\mathrm{E}}\left(
	\frac{\psi^\mathrm{E}}{H^\mathrm{E}}
	\right)
	\nonumber\\&=
	\mathcal{A}^\mathrm{J}_\psi
	+\frac{f' \delta\varphi}{2 f}
	-\frac{1}{\sqrt{f}}\frac{d}{dt^\mathrm{J}} \left[
	\left(
	\frac{1}{\sqrt{f}} H^\mathrm{J}
	+\frac{f' \dot{\varphi}}{2 f^{3/2}}
	\right)^{-1}
	\left(
	\frac{f' \delta\varphi}{2 f}
	\right)
	\right]
	\,.
\end{align}
Using the solution \eqref{eqn:sol-Apsi-DF}, we find $\dot{\mathcal{A}}^\mathrm{J}_\psi \approx 2 H^\mathrm{J} \epsilon^\mathrm{J}_1 \mathcal{A}^\mathrm{J}_\psi$, where $\epsilon^\mathrm{J}_1$ is the first Hubble slow-roll parameter in the Jordan frame\footnote{
	Note that $\lambda$ corresponds to $\beta$ in Ref.~\cite{DeFelice:2023psw}.
},
\begin{align}
	\epsilon^\mathrm{J}_1 \approx
	\frac{1}{36 \lambda (H^\mathrm{J})^2}
	\,,
\end{align}
which holds under the slow-roll approximation; see also Ref.~\cite{DeFelice:2023psw}.
We may thus express the difference, $\Delta\mathcal{A}_\psi \equiv \mathcal{A}^\mathrm{E}_\psi - \mathcal{A}^\mathrm{J}_\psi$, as follows:
\begin{align}
	\Delta\mathcal{A}_\psi \approx
	-\mathcal{A}^\mathrm{J}_\psi
	+\frac{3 \epsilon^\mathrm{J}_1}{2} \mathcal{A}^\mathrm{J}_\psi
	\,,
\end{align}
up to the first order in the Jordan-frame slow-roll parameters, together with the $\lambda (H^\mathrm{J})^2 \gg 1$ limit.
We thus see that
\begin{align}
	\mathcal{A}^\mathrm{E}_\psi \approx
	\frac{3 \epsilon^\mathrm{J}_1}{2} \mathcal{A}^\mathrm{J}_\psi
	\,.
\end{align}
Without taking into account the slow-roll contribution, one would observe $\mathcal{A}^\mathrm{E}_\psi \approx 0$.

Let us now examine the growth rate of $\mathcal{A}^\mathrm{E}_\psi$ in the Jordan frame. Taking the time derivative, we obtain
\begin{align}
	\frac{d \ln \mathcal{A}^\mathrm{E}_\psi}{d t^\mathrm{J}} \approx
	H (2 \epsilon^\mathrm{J}_1 + \epsilon^\mathrm{J}_2)
	\,,
\end{align}
up to the leading order in the slow-roll parameters and in the $\lambda (H^\mathrm{J})^2 \gg 1$ limit, where $\epsilon^\mathrm{J}_2 \equiv \dot{\epsilon}^\mathrm{J}_1/(H^\mathrm{J} \epsilon^\mathrm{J}_1)$. Using $H^\mathrm{E} dt^\mathrm{E} = H^\mathrm{J} dt^\mathrm{J} + \mathcal{O}(\epsilon^\mathrm{J}_1)$, we obtain the growth rate of $\mathcal{A}^\mathrm{E}_\psi$ in the Einstein frame as
\begin{align}
	\frac{d \ln \mathcal{A}^\mathrm{E}_\psi}{d t^\mathrm{E}} &\approx
	H^\mathrm{E} (2 \epsilon^\mathrm{J}_1 + \epsilon^\mathrm{J}_2)
	\,.
\end{align}
Since $\epsilon^\mathrm{J}_2 \approx 2 \epsilon^\mathrm{J}_1$ under the same slow-roll approximation and $\epsilon^\mathrm{J}_1 \approx \epsilon^\mathrm{E}_2 / 4$, we get
\begin{align}
	\frac{d \ln \mathcal{A}^\mathrm{E}_\psi}{d t^\mathrm{E}} &\approx
	H^\mathrm{E} \epsilon^\mathrm{E}_2
	\,,
\end{align}
recovering the behaviour we observed in the Einstein-frame analysis, namely Eq.~\eqref{eqn:sol-Apsi-NLO}.

Another way to show the equivalence is by showing the equivalence of the perturbed equations of motion. Under the conformal transformation with the choice of
\begin{align}
	E^\mathrm{E} = 0
	\,,\quad
	\psi^\mathrm{E} = \frac{f' \delta\varphi}{2 f}
	\,,
\end{align}
which corresponds to the Jordan-frame flat gauge, the Einstein-frame perturbed equations of motion become\footnote{
	Note that $\omega$ corresponds to $2\alpha$ in Ref.~\cite{DeFelice:2023psw}.
}
\begin{align}
	0 &=
	-108 \lambda (H^\mathrm{J})^2 \dot{(H^\mathrm{J})} A^\mathrm{J}
	+18 \lambda (\dot{H}^\mathrm{J})^2 A^\mathrm{J}
	-36 \lambda H^\mathrm{J} \ddot{H}^\mathrm{J} A^\mathrm{J}
	-54 \lambda (H^\mathrm{J})^3 \dot{A}^\mathrm{J}
	-36 \lambda H^\mathrm{J} \dot{H}^\mathrm{J} \dot{A}^\mathrm{J}
	\nonumber\\&\quad
	-18 \lambda (H^\mathrm{J})^2 \ddot{A}^\mathrm{J}
	+H^\mathrm{J} \frac{k^2}{a^\mathrm{J}} B^\mathrm{J}
	-12 \lambda (H^\mathrm{J})^3 \frac{k^2}{a^\mathrm{J}} B^\mathrm{J}
	+36 \lambda H^\mathrm{J} \dot{H}^\mathrm{J} \frac{k^2}{a^\mathrm{J}} B^\mathrm{J}
	+6 \lambda \ddot{H}^\mathrm{J} \frac{k^2}{a^\mathrm{J}} B^\mathrm{J}
	\nonumber\\&\quad
	+6 \lambda (H^\mathrm{J})^2 \frac{k^2}{a^\mathrm{J}} \dot{B}^\mathrm{J}
	-6 \lambda \dot{H}^\mathrm{J} \frac{k^2}{a^\mathrm{J}} \dot{B}^\mathrm{J}
	+6 \lambda H^\mathrm{J} \frac{k^2}{a^\mathrm{J}} \ddot{B}^\mathrm{J}
	-42 \lambda (H^\mathrm{J})^2 \frac{k^2}{(a^\mathrm{J})^2} A^\mathrm{J}
	-18 \lambda \dot{H}^\mathrm{J} \frac{k^2}{(a^\mathrm{J})^2} A^\mathrm{J}
	\nonumber\\&\quad
	+6 \lambda H^\mathrm{J} \frac{k^4}{(a^\mathrm{J})^3} B^\mathrm{J}
	+2 \lambda \frac{k^4}{(a^\mathrm{J})^3} \dot{B}^\mathrm{J}
	-\frac{\omega}{3} \frac{k^4}{(a^\mathrm{J})^3} \dot{B}^\mathrm{J}
	+2 \lambda \frac{k^4}{(a^\mathrm{J})^4} A^\mathrm{J}
	-\frac{\omega}{3} \frac{k^4}{(a^\mathrm{J})^4} A^\mathrm{J}
	\,,\label{eqn:JFperteom1}
\end{align}
and
\begin{align}
	0 &=
	-3 a^\mathrm{J} (H^\mathrm{J})^2 B^\mathrm{J}
	-2 a^\mathrm{J} \dot{H}^\mathrm{J} B^\mathrm{J}
	-108 \lambda a^\mathrm{J} (H^\mathrm{J})^2 \dot{H}^\mathrm{J} B^\mathrm{J}
	-54 \lambda a^\mathrm{J} (\dot{H}^\mathrm{J})^2 B^\mathrm{J}
	-72 \lambda a^\mathrm{J} H^\mathrm{J} \ddot{H}^\mathrm{J} B^\mathrm{J}
	\nonumber\\&\quad
	-12 \lambda a^\mathrm{J} \dddot{H}^\mathrm{J} B^\mathrm{J}
	+2 H^\mathrm{J} A^\mathrm{J}
	+144 \lambda H^\mathrm{J} \dot{H}^\mathrm{J} A^\mathrm{J}
	+36 \lambda \ddot{H}^\mathrm{J} A^\mathrm{J}
	+36 \lambda (H^\mathrm{J})^2 \dot{A}^\mathrm{J}
	+36 \lambda \dot{H}^\mathrm{J} \dot{A}^\mathrm{J}
	\nonumber\\&\quad
	+12 \lambda H^\mathrm{J} \ddot{A}^\mathrm{J}
	+24 \lambda (H^\mathrm{J})^2 \frac{k^2}{a^\mathrm{J}} B^\mathrm{J}
	-12 \lambda \dot{H}^\mathrm{J} \frac{k^2}{a^\mathrm{J}} B^\mathrm{J}
	-4 \lambda H^\mathrm{J} \frac{k^2}{a^\mathrm{J}} \dot{B}^\mathrm{J}
	+\frac{2 \omega}{3} H^\mathrm{J} \frac{k^2}{a^\mathrm{J}} \dot{B}^\mathrm{J}
	\nonumber\\&\quad
	-4 \lambda \frac{k^2}{a^\mathrm{J}} \ddot{B}^\mathrm{J}
	+\frac{2 \omega}{3} \frac{k^2}{a^\mathrm{J}} \ddot{B}^\mathrm{J}
	+12 \lambda H^\mathrm{J} \frac{k^2}{(a^\mathrm{J})^2} A^\mathrm{J}
	-4 \lambda \frac{k^2}{(a^\mathrm{J})^2} \dot{A}^\mathrm{J}
	+\frac{2\omega}{3} \frac{k^2}{(a^\mathrm{J})^2} \dot{A}^\mathrm{J}
	\,,\label{eqn:JFperteom2}
\end{align}
where we have utilised the Jordan-frame background equations of motion \cite{DeFelice:2023psw},
\begin{align}
	\dddot{H}^\mathrm{J} &=
	-\frac{72 H^\mathrm{J} \ddot{H}^\mathrm{J} \lambda + 2 \dot{H}^\mathrm{J} (1+27\dot{H}^\mathrm{J} \lambda) + 3 (H^\mathrm{J})^2 (1+36\dot{H}^\mathrm{J} \lambda)}{12 \lambda}
	\,,\\
	\ddot{H}^\mathrm{J} &= 
	-\frac{(H^\mathrm{J})^2 + 36 (H^\mathrm{J})^2 \dot{H}^\mathrm{J} \lambda - 6 (\dot{H}^\mathrm{J})^2 \lambda}{12 H^\mathrm{J} \lambda}
	\,.
\end{align}
We can eliminate $B^\mathrm{J}$ by manipulating Eqs.~\eqref{eqn:JFperteom1} and \eqref{eqn:JFperteom2}, obtaining the evolution equation for $A^\mathrm{J}$, which is identically $\mathcal{A}^\mathrm{J}_\psi$ as we are in the Jordan-frame flat gauge. The resultant equation is given by
\begin{align}
	0 &=
	\ddddot{\mathcal{A}}^\mathrm{J}_\psi
	+\tau_1 H^\mathrm{J} \dddot{\mathcal{A}}^\mathrm{J}_\psi
	+\tau_2 (H^\mathrm{J})^2 \ddot{\mathcal{A}}^\mathrm{J}_\psi
	+\tau_3 (H^\mathrm{J})^3 \dot{\mathcal{A}}^\mathrm{J}_\psi
	+\tau_4 (H^\mathrm{J})^4 \mathcal{A}^\mathrm{J}_\psi
	\,.
\end{align}
The coefficients $\tau_i$ are, in the subhorizon limit, given by
\begin{align}
	\tau_1 &\approx
	10
	\,,\quad
	\tau_2 \approx
	2\left(\frac{k}{a^\mathrm{J} H^\mathrm{J}}\right)^2
	\,,\quad
	\tau_3 \approx
	6\left(\frac{k}{a^\mathrm{J} H^\mathrm{J}}\right)^2
	\,,\quad
	\tau_4 \approx
	\left(\frac{k}{a^\mathrm{J} H^\mathrm{J}}\right)^4
	\,.
\end{align}
In the superhorizon limit, if we further impose the condition $\lambda (H^\mathrm{J})^2 \gg 1$, the coefficients $\tau_i$ are given by
\begin{align}
	\tau_1 &\approx 6
	\,,\quad
	\tau_2 \approx 11 + \frac{24\lambda}{\omega}
	\,,\quad
	\tau_3 \approx 6 + \frac{72\lambda}{\omega}
	\,,\quad
	\tau_4 \approx -\frac{\omega+12\lambda}{3H^2\omega\lambda}
	\,.
\end{align}
These results are identical to those reported in Ref.~\cite{DeFelice:2023psw}.
We stress that the same analysis can be performed in other gauge choices. We have checked, following the aforementioned procedure, the equivalence between the Einstein-frame analysis and the Jordan-frame analysis in the other gauge choices.


\bibliographystyle{apsrev4-2}
\bibliography{main}
\end{document}